\documentclass[amssymb,amsmath,aps,showpacs,floatfix,nofootinbib,12pt]{article}

\evensidemargin=\oddsidemargin

\usepackage{graphicx,color,amsmath,amssymb,color,appendix}
\usepackage{epstopdf}
\usepackage{multirow,comment}
\usepackage{colortbl}
\usepackage{longtable}
\usepackage{subcaption}
\usepackage{slashed}
\usepackage{isodateo}
\usepackage[CJKbookmarks=true,bookmarksnumbered=true,bookmarksopen=true]{hyperref}
\hypersetup{colorlinks,%
              linkcolor=blue,
              citecolor=blue,
              urlcolor=blue}

\usepackage[dotinlabels]{titletoc}
\usepackage{titlesec}
\usepackage{ulem}

\definecolor{LightBlue}{cmyk}{0.25,0,0,0}

\newcolumntype{C}[1]{>{\centering\arraybackslash}p{#1}}

\newcommand{\be}{\begin{equation}}
\newcommand{\ee}{\end{equation}}
\newcommand{\bea}{\begin{eqnarray}}
\newcommand{\eea}{\end{eqnarray}}
\newcommand{\bp}{\begin{pmatrix}}
\newcommand{\ep}{\end{pmatrix}}

\newcommand{\ra}{\rangle }
\newcommand{\la}{\langle }

\newcommand{\fr}[2]{\mbox{$\frac{\,{#1}\,}{#2}$}}
\renewcommand{\rm}{\mathrm}
\newcommand{\GeV}{\text{GeV}}
\newcommand{\MeV}{\text{MeV}}
\newcommand{\keV}{\text{keV}}
\newcommand{\eV}{\text{eV}}
\newcommand{\Mpc}{\text{Mpc}}
\newcommand{\SEC}{\text{s}}
\newcommand{\km}{\text{km}}

\newcommand{\beq}{\begin{equation}}
\newcommand{\eeq}{\end{equation}}
\newcommand{\bq}{\begin{equation}}
\newcommand{\eq}{\end{equation}}
\newcommand{\ba}{\begin{array}}
\newcommand{\ea}{\end{array}}
\newcommand{\beqa}{\begin{eqnarray}}
\newcommand{\eeqa}{\end{eqnarray}}
\newcommand{\beqs}{\begin{subequations}}
\newcommand{\eeqs}{\end{subequations}}

\def\nn{\nonumber}

\def\dis{\displaystyle}

\def\({\left(}
\def\){\right)}
\newcommand{\calO}{{\cal O} }

\def\ZZ{\mathbb{Z}_2^{}}
\def\LRto{\leftrightarrow}
\def\eq{\text{eq}}
\def\Hz{H_0^{}}

\begin{document}

 \thispagestyle{empty}
 \renewcommand{\thefootnote}{\fnsymbol{footnote}}
 \setcounter{footnote}{0}
 \titlelabel{\thetitle.\quad \hspace{-0.8em}}
\titlecontents{section}
              [1.5em]
              {\vspace{2.5mm} \large}
              {\contentslabel{1em}}
              {\hspace*{-1em}}
              {\titlerule*[.5pc]{.}\contentspage}
\titlecontents{subsection}
              [3.5em]
              {\vspace{2mm}}
              {\contentslabel{1.8em}}
              {\hspace*{.3em}}
              {\titlerule*[.5pc]{.}\contentspage}
\titlecontents{subsubsection}
              [5.5em]
              {\vspace{2mm}}
              {\contentslabel{2.5em}}
              {\hspace*{.3em}}
              {\titlerule*[.5pc]{.}\contentspage}
\titlecontents{appendix}
              [1.5em]
              {\vspace{2.5mm} \large\bf}
              {\contentslabel{1em}}
              {\hspace*{-1em}}
              {\titlerule*[.5pc]{.}\contentspage}


\begin{center}
{\Large\bf  Resolving Hubble Tension by \\[2mm]
Self-Interacting Neutrinos with Dirac Seesaw
}

\vspace*{8mm}

{\sc Hong-Jian He},$^{a,b,c}$\footnote{Email: hjhe@sjtu.edu.cn}~
{\sc Yin-Zhe Ma},$^{d,e}$\footnote{Email: ma@ukzn.ac.za}~
{\sc Jiaming Zheng}\,$^{a}$\footnote{Email: zhengjm3@sjtu.edu.cn}

\vspace*{4mm}

$^a$\,Tsung-Dao Lee Institute $\&$ School of Physics and Astronomy, \\
Shanghai Key Laboratory for Particle Physics and Cosmology,\\
Shanghai Jiao Tong University, Shanghai 200240, China
\\[2mm]
$^b$\,Institute of Modern Physics and Department of Physics, \\
Tsinghua University, Beijing 100084, China
\\[2mm]
$^{c}$\,Center for High Energy Physics, Peking University, Beijing 100871, China
\\[2mm]
$^d$\,School of Chemistry and Physics, University of KwaZulu-Natal,\\
Westville Campus, Private Bag X54001, Durban, 4000, South Africa
\\[2mm]
$^e$\,NAOC-UKZN Computational Astrophysics Centre (NUCAC),\\
University of KwaZulu-Natal, Durban, 4000, South Africa

\end{center}

\vspace*{5mm}

\begin{center}
{\large\bf Abstract}
\end{center}
\vspace*{-3mm}
\noindent
Self-interacting neutrinos that begin to free-stream at close to matter-radiation equality can reduce the physical size of photon sound horizon at last scattering surface. This mechanism can be the reason why standard $\Lambda$CDM cosmology sees a lower value of the Hubble constant than local measurements from distance ladder.
We propose a new realization of self-interacting Dirac neutrinos (SID$\nu$) with
light-dark-photon mediator for a viable interaction mechanism.
Our model is UV completed by a Dirac seesaw with
anomaly-free dark $U(1)_X^{}$ gauge group which charges the right-handed neutrinos. This model naturally generates small masses for Dirac neutrinos and induces self-scattering of right-handed neutrinos. The scattering with left-handed neutrinos is suppressed by a chirality-flip mass insertion when the neutrino energy is much larger than its mass. The resultant neutrino self-scattering is not operative for $E_{\nu}\gtrsim O(\text{keV})$, which avoids the cosmological and laboratory constraints. By evolving Boltzmann equations for left- and right-handed neutrino number densities, we show that about $2/3$ of the left-handed neutrinos are converted into right-handed neutrinos in a short epoch between the Big-Bang Nucleosynthesis and the recombination, and interact with each other efficiently afterwards. The resultant neutrino non-free-streaming is the crucial ingredient to shrink down the comoving sound horizon at drag epoch, which can reconcile the Hubble tension between early and late time measurements.
\\[5mm]
JCAP (2020), in Press $[$\,arXiv:2003.12057\,$]$.

\newpage
\renewcommand{\thefootnote}{\arabic{footnote}}
\setcounter{footnote}{0}
\setcounter{page}{2}

\baselineskip 16.5pt
\tableofcontents

\setcounter{footnote}{0}
\renewcommand{\thefootnote}{\arabic{footnote}}
\numberwithin{equation}{section}

\baselineskip 17.5pt

\vspace*{6mm}
\section{Introduction}
\label{sec:1}
\label{sec:intro}

The discrepancy between measurements of the Hubble constant $\Hz$ from the observations of the early Universe and from the late time observations poses a severe challenge to the conventional $\Lambda$-Cold Dark Matter ($\Lambda$CDM) cosmology\,\cite{Bernal:2016gxb,Verde:2019ivm}.
In particular, {\it Planck} space telescope measures $\Hz$ from
the cosmic microwave background (CMB) and gives
$\,\Hz\!=67.4 \pm 0.5\,{\km\,\SEC^{-1} \Mpc^{-1}}$ \cite{Aghanim:2018eyx},\,
with a precision better than 1\%.
This is compatible with the independent result from
Baryon Acoustic Oscillation (BAO) \& Dark Energy Survey (DES) \&
Big Bang Nucleosynthesis data~\cite{Abbott:2017smn}, which gives
$\,H_0^{}\!=67.4^{+1.1}_{-1.2}\,{\km\,\SEC^{-1}\Mpc^{-1}}$.\,
In contrast, the distance ladder measurement (SH0ES) by using Type-Ia supernovae calibrated
by Cepheid favors a larger Hubble constant,
$\,H_0^{}\!=\!74.0\pm 1.4\,{\km\,\SEC^{-1}\Mpc^{-1}}$~\cite{Riess:2019cxk}.
This result is consistent with another completely
independent measurement of the strong lensing time-delay effect.
By measuring six distant quasar time-delays, the H0LiCOW team determines $\,H_0^{}\!=73.3^{+1.7}_{-1.8}\,{\km\,\SEC^{-1} \Mpc^{-1}}$~\cite{Wong:2019kwg}.
This solidifies the discrepancy between high-redshift measurements and local measurements.
Although the systematic uncertainty of the distance ladder measurement is
under debate\,\cite{Rameez:2019wdt} and the distance ladder calibrated by tip of the red giant branch gives
$\,H_0^{}\!=69.8\pm0.8\pm\!1.7\,{\km\,\SEC^{-1} \Mpc^{-1}}$~\cite{Freedman:2019jwv}, consistent with early time observations,
a recent survey on various $\Hz$ measurements concludes
that the $\Hz$ discrepancy between early and late Universe observations
ranges from $4\sigma$ to $6\sigma$, and is robust to the exclusion of any one method, team or source\,\cite{Verde:2019ivm}.

\vspace*{1mm}

A physically attractive resolution to the Hubble tension is the scenario of self-interacting neutrinos\,\cite{Cyr-Racine:2013jua,Kreisch:2019yzn,Forastieri:2019cuf,Ghosh:2019tab,Blinov:2019gcj,Blinov:2020hmc},
but its viable realization was found to be highly challenging\,\cite{Blinov:2019gcj}.
In this scenario, the onset of neutrino free-streaming is delayed in the early universe, and the resultant phase shift and amplification of acoustic peaks in the CMB power spectrum can be compensated by shifts of other cosmological parameters\,\cite{Cyr-Racine:2013jua,Kreisch:2019yzn,Oldengott:2017fhy,Lancaster:2017ksf}.
In particular, Refs.\,\cite{Kreisch:2019yzn,Blinov:2019gcj} found that if the active neutrinos self-interact through an effective vertex
\beqa
\label{eq:Geff-4nu}
{\cal L}_\text{eff}^{}\,=\, G_\text{eff}^{}\,\bar{\nu}{\nu}\bar{\nu}{\nu}\,,
\eeqa
a larger Hubble constant $H_0^{}\!=72.3\pm 1.4\,{\km\,\SEC^{-1} \Mpc^{-1}}$ (with
$\Delta N_\text{eff}^{}\approx 1$) can be accommodated by the CMB observation
for the ``strongly interacting'' regime and ``moderately interacting'' regime
with
$\,\text{log}_{10}^{}(G_\text{eff}^{} \MeV^2 )\!=\! -1.35^{+0.12}_{-0.066}$\,
and $-3.90^{+1.0}_{-0.93}$, respectively,
from the Planck\,TT\,+\,lens\,+\,BAO\,+\,$H_0$\, datasets.
Ref.\,\cite{Kreisch:2019yzn} considered an effective interaction
of neutrino mass-eigenstates in the form
\beqa
\label{eq:nunu-phi}
{\cal L}\,=\,g_{ij}^{}\bar{\nu}_i^{}{\nu}_j^{}\varphi\,,
\eeqa
and found that the Hubble tension can be evaded with
$G_{\text{eff}}^{}\!\equiv\! g^2/m_\varphi^2\,
\!=\! (10^{-1}\!-\!10^{-4})\,
\text{MeV}^{-2}$ and $\Delta N_{\text{eff}}^{}\!\approx\! 1$\,.
However, Ref.\,\cite{Forastieri:2019cuf} found that neutrino self-interactions induced by a very light or massless mediator cannot resolve the Hubble tension. Ref.\,\cite{Ghosh:2019tab} considered a possibility that the neutrino free-streaming is impeded by the ``dark neutrino interaction'' between neutrinos and the dark matter, and found that the phase shift of non-free-streaming neutrinos alone can raise the CMB determined Hubble constant to $H_0^{}\!=\!69.39^{+0.69}_{-0.68}\,\text{km\,s}^{-1}\text{Mpc}^{-1}$
without additional $\Delta N_{\text{eff}}^{}$\,.
These suggest that the Hubble tension could be resolved if the
neutrino free-streaming does not turn on before $\,T\!\sim\!10$\,eV
when the modes relevant to the observed CMB power spectrum enter the horizon.
But the neutrino self-interactions \eqref{eq:Geff-4nu} and \eqref{eq:nunu-phi}
are {\it not gauge-invariant.}
It was found that a UV completion is highly constrained
and almost excluded by cosmological observations such as
the Big Bang Nucleosynthesis
(BBN)\,\cite{Blinov:2019gcj,Cyburt:2015mya,Berlin:2019pbq,Grohs:2020xxd},
or by laboratory bounds such as meson decays\,\cite{Blinov:2019gcj,Pasquini:2015fjv,Britton:1992xv,Artamonov:2014urb}.
Furthermore, the light neutrinos have to be Majorana type, the neutrino self-interaction
needs to be flavor-dependent, and the UV-completion model requires
a nonminimal mechanism to simultaneously generate
neutrino masses and appreciable self-interactions\,\cite{Blinov:2019gcj}.
Some other different attempts to alleviate the Hubble tension with neutrino physics
appeared in \cite{Rodrigues:2020dod,Escudero:2019gvw,Sakstein:2019fmf,Gehrlein:2019iwl}.

\vspace*{1mm}

In this work, we propose a physically attractive model of
self-interacting Dirac neutrinos (SID$\nu$)
with light-dark-photon mediator to delay the neutrino free-streaming time-scale,
and thus shrink the comoving sound horizon at the last scattering surface ($r_{\ast}^{}$)
without drastically affecting the projected Silk damping scale ($\ell_d^{}$).
Such modification of the early time physics will result in an increased Hubble rate inferred by the CMB measurement.
Our new model is UV-completed by a Dirac seesaw with an
anomaly-free dark $U(1)_X^{}$ gauge group which charges the right-handed neutrinos
and is spontaneously broken.
This mechanism naturally generates small masses for Dirac neutrinos and simultaneously induces
self-interacting scattering of right-handed neutrinos.
Thus, different from the previous literature\,\cite{Kreisch:2019yzn,Forastieri:2019cuf,Ghosh:2019tab,Blinov:2019gcj},
our model has the right-handed neutrinos (rather than the left-handed ones)
interact with the dark photon $X^\mu$ (rather than a scalar $\varphi$)
at an energy scale of $O(\MeV)$.\,
The dark photon $X^\mu$ serves as the mediator
of the hidden neutrino interaction, which is a key ingredient of our scenario.
In the early Universe, only left-handed neutrinos are produced abundantly
from the thermal bath of the standard model (SM) particles by electroweak interactions.
The scattering amplitude of neutrinos through the dark photon exchange
is suppressed by a chirality-flip (mass-insertion) factor $m_\nu^{}/E_\nu^{}$
for each left-handed neutrino participating in the scattering,
where $E_\nu^{}$ and $m_\nu^{}$ are the neutrino energy and mass, respectively.
Hence, the production of right-handed neutrinos and the mediator particles
from left-handed neutrino scattering is suppressed at high temperature,
so it is free from cosmological constraints such as the strong BBN bound.
As the temperature decreases, the chirality-flip factor becomes larger and
has less suppression. At the temperature
$\,T_{\rm c}^{}\!\ll\! O(\MeV)$,\,
the small amount of right-handed neutrinos produced out-of-equilibrium start to scatter effectively with left-handed neutrinos,
and trigger a rapid conversion of left-handed neutrinos to the right-handed ones.
Eventually, the cosmic neutrino relics are composed of both left-handed and
right-handed neutrinos which scatter efficiently with each other until
the decoupling of the dark photon interaction
at which the neutrinos begin to free-stream.
In this way, we build up a consistent and novel realization
of the self-interacting neutrino scenario as a resolution to the Hubble tension,
which overcomes all the difficulties in the previous proposal\,\cite{Kreisch:2019yzn}.
Moreover, our model naturally generates the small Dirac neutrino masses
and does not require any special flavor structure of the neutrino interaction
to evade all the existing cosmological
and laboratory constraints\,\cite{Blinov:2019gcj}.

\vspace*{1mm}

The rest of this paper is organized as follows.
In Section\,\ref{sec:Int_Dirac_nu}, we propose a new realization of
Dirac neutrino seesaw as the UV completion of self-interacting neutrinos
in the early Universe.
In Section\,\ref{sec:RHnu_evol}, we analyze qualitatively the evolution of
the right-handed neutrinos in the early Universe
and the condition to delay the free-streaming,
while evading the cosmological and laboratory constraints.
In Section\,\ref{sec:evolution}, we perform numerical analysis
to evolve the neutrino energy density by Boltzmann equations
as an explicit demonstration of the physical picture described
in Section\,\ref{sec:RHnu_evol}. Finally, we conclude in Section\,\ref{sec:conclusion}.
We present the technical details in Appendices\,\ref{A:A} and \ref{A:B}.

\vspace*{2mm}
\section{\large Interacting Dirac Neutrinos from Dirac Seesaw}
\label{sec:Int_Dirac_nu}
\label{sec:2}

In this section, we show that the neutrino
self-interaction can be naturally realized in a new Dirac seesaw model of neutrinos
with a dark $U(1)_X^{}$ gauge group.
The Dirac seesaw was proposed\,\cite{Gu:2006dc}
to generate small Dirac masses for light neutrinos.
Its crucial part contains the right-handed neutrinos with charge $-1/2$
under a hidden dark ${U(1)}_X^{}$ gauge group.
This ${U(1)}_X^{}$ is spontaneously broken by a weak singlet scalar $S$
at the TeV scale (or somewhat below) which has a ${U(1)}_X^{}$ charge $1/2$.\,
This can generate a gauge-invariant dimension-5 effective operator at the weak scale
for the Dirac neutrino mass generation,
%
$\,\calO_5^{}\!=\!\frac{1}{\,\Lambda\,}\,\bar{L}HS\nu_R^{}$\,,
%
where $\Lambda$ is a high energy cutoff scale,
$L$ the left-handed lepton doublet and $H$ the SM Higgs doublet.
So the light neutrinos acquire small Dirac masses
$\,m_\nu^{}\!\sim\!\left<H\right>\!\left<S\right>\!/\Lambda$\,.\,

\vspace*{1mm}

\begin{table}[t]
	\begin{center}
		\renewcommand{\arraystretch}{1.3} 
		\begin{tabular}{c||c|c|c|c|c|c|c}
			\hline\hline
			Groups  & $L_j^{}$  &  $H$ & $\Phi_1^{}$ & $\Phi_2^{}$ & $S$ & $R_{1j}^{}$ & $R_{2j}^{}$
			\\
			\hline\hline
			$~{SU(2)}_L^{}~$ & {\bf 2} & {\bf 2} & {\bf 2} & {\bf 2} & {\bf 1} &{\bf 1}&{\bf 1}
			\\ \hline
			${U(1)}_{Y}$ &  ~$-\frac{1}{2}$~ & ~$-\frac{1}{2}$~ &
			~$-\frac{1}{2}$~  & ~$-\frac{1}{2}$~  & $0$ & $0$& $0$
			\\ \hline
			${U(1)}_{X}$ & $0$ & $0$ & $\frac{1}{2}$ &
			$-\frac{1}{2}$ & ~~$\frac{1}{2}$~~  & ~$-\frac{1}{2}$~ &
			~~$\frac{1}{2}$~~
			\\
			\hline\hline
		\end{tabular}
		\renewcommand{\arraystretch}{1} 
	\end{center}
	\vspace*{-3mm}
	\caption{\small
		Assignments for the Dirac seesaw model under the extended electroweak gauge group
		$SU(2)_L^{}\!\otimes\! U(1)_Y^{}\!\otimes\!U(1)_X^{}$. Here
		$\,j\,(=\!1,2,3)$  denotes the index of fermion families.
		\label{tab:z2rep}
		\label{tab:1}
	}
\end{table}

For this study, we propose a new realization of the Dirac seesaw mechanism
with an anomaly-free dark $U(1)_X^{}$ gauge group, a conserved lepton number
at the classical level and an exact $\mathbb{Z}_2^{}$
symmetry. This naturally extends the previous simple model\,\cite{Gu:2006dc}
which was not UV-completed for anomaly cancellation.
We present this model in Table\,\ref{tab:z2rep},
where $\Phi_1^{}$ and $\Phi_2^{}$ are
two new heavy Higgs doublets with mass $M_\Phi^{}\!=O(10^9\GeV)$.\,
The light singlet scalar $S$ acquires a vacuum expectation value (VEV)
of $O(\text{MeV})$ and spontaneously breaks $U(1)_X^{}$ gauge group,
leading to a dark photon of mass around
$O(\text{keV})$.\,
$R_{1j}^{}$ and $R_{2j}^{}$ are two right-handed Dirac fermions which carry opposite
${U(1)}_X^{}$ charges to cancel the gauge anomaly,
where $j\,(=\!1,2,3)$ denotes the fermion family index.
As we will show shortly, the combination $R_{1j}^{}\!+\!R_{2j}^{}$ just gives
the right-handed neutrinos $\nu_{Rj}^{}$.\,
Assigning $R_{1j}^{}$ and $R_{2j}^{}$ to have
the same lepton number as $L_j^{}$\,,\,
we can write down the lepton number conserving Lagrangian terms
relevant to the Dirac seesaw,
\beqa
\Delta{\cal L} &\supset&
-y_{ij}^{} \bar{L}_i^{} \!\( \Phi_1^{}R_{1j}^{} \!
+ \Phi_2^{} R_{2j}^{} \)
+ M_3^{} ( S \Phi_1^\dagger \!+ S^* \Phi_2^\dagger )H
+ \text{h.c.}
\nn\\[1mm]
&& -M_\Phi^2 \!\(|\Phi_1^{}|^2 \!+ |\Phi_2^{}|^2\),
\label{eq:L1}
\eeqa
where $i,j=1,2,3$ are the family indices, and the trilinear scalar coupling may
be around the $\Phi$ mass scale,
$\,M_3^{} =O(M_\Phi^{})$.\,
The Lagrangian is invariant under the following $\mathbb{Z}_2^{}$ symmetry,
\beqa
\mathbb{Z}_2^{}\!:
\qquad
B^\mu \leftrightarrow  B^\mu ,
\quad
X^\mu \leftrightarrow -X^\mu ,
\quad
\Phi_{1}^{}\leftrightarrow \Phi_{2}^{},
\quad
S\leftrightarrow S^* ,
\quad
R_{1j}^{}\leftrightarrow R_{2j}^{},
\quad
\eeqa
where $B^\mu$ and $X^\mu$ are gauge bosons of $U(1)_Y^{}$ and $U(1)_X^{}$, respectively.
The above $\ZZ$ assignments can be re-expressed as follows,
\beq
\ba{rccccc}
\text{Fields}:~ & ~B^\mu~ & ~X^\mu~ &
~\Phi_1^{}\pm\Phi_2^{}~  & ~S\pm S^*~ & ~R_{1j}^{}\pm R_{2j}^{}~
\\[1.5mm]
\ZZ :~ & +~ & -~ & \pm & \pm\, & \pm
\ea
\eeq
This $\ZZ$ symmetry forbids the kinetic mixing between $X^\mu$ and $B^\mu$
to all loop orders, and thus can evade possible astrophysical constraints
on the light-dark-photons\,\cite{DPbound}.
Since $M_\Phi\gg m_S^{}, m_H^{}$,
we can integrate out the heavy fields $\Phi_1$ and $\Phi_2$
by using their equations of motions,
\beqs
\vspace*{-4mm}
\beqa
{\Phi_1^{}} &\!=\!&
\frac{M_3^{}}{M_\Phi^2} H S
-\frac{y_{ij}^*}{M_\Phi^2} \bar{R}_{1j} L_i + \cdots ,
\\
{\Phi_2^{}}  &\!=\!&
\frac{M_3^{}}{M_\Phi^2} H S^*
-\frac{y_{ij}^*}{M_\Phi^2} \bar{R}_{2j} L_i + \cdots .
\eeqa
\eeqs
With this we can deduce the following effective Lagrangian
from Eq.\eqref{eq:L1},
\beqa
\Delta{\cal L}\,=\,
-\frac{\,y_{ij}^{}M_3^{}\,}{M_\phi^2}
\bar{L}_i^{} H
{\left( S R_{1j}^{}\! + S^*\!R_{2j}^{}\right)}
\!+ \text{h.c.}
+\cdots
\label{eq:z2yukawa}
\eeqa
Integrating out the heavy Higgs doublets $\Phi_{1,2}^{}$
will also induce a correction to the quartic coupling
$\,\sim\!(M_3^2/M_\Phi^2)|S|^2|H|^2$,\, which is added to the original tree-level
Higgs portal term $|S|^2|H|^2$ with a total coupling $\lambda_{SH}^{}$.\,
For the current setup, we set the coupling
$\,\lambda_{SH}^{}\!=0$\, at tree level.
With this choice,
the $|S|^2|H|^2$ vertex will remain suppressed at loop levels at low energy scales.
The loop contribution to this vertex from Eq.\eqref{eq:z2yukawa} is suppressed by
$\left(\!\frac{M_3^{}}{\,M_\phi^2\,}\!\right)^{\!\!2}$.
We also note that
including the graviton-exchange contribution between $S$ and $H$
could only induce a nonlocal interaction between $|S|^2$ and $|H|^2$,
which is suppressed by the Planck mass factor $M_{\text{Pl}}^{-2}$
and thus negligible.
Assuming a reheating temperature much less than $M_\phi^{}$\,,
these corrections are irrelevant to the evolution of the hot plasma.
Hence, we can avoid the production of $\,S\,$
through the Higgs portal coupling in the early Universe,
while maintaining a light scalar $\,S\,$ with
$\,m_S^{}\!\ll\! M_{h}^{}$\,,\,
where $M_{h}^{}\!\simeq\! 125\,$GeV
is the SM Higgs boson mass.

\vspace*{1mm}

After $S$ and $H$ develop the VEVs $\,\la S \ra \!=v_s^{}/\!\sqrt{2}$\, and
$\,\la H \ra \!=\!(v_h^{}/\!\sqrt{2},\,0)$,\,
we find that the neutrinos acquire the following Dirac mass term,
\beq
\ba{rl}
{\cal L}_\nu^{}  &\!=\, -m_{\nu ij}^{}
\bar{\nu}_{Li}^{} \nu_{Rj}^{} + \text{h.c.}
\,,
\\[2mm]
m_{\nu ij}^{} &\!=\,
\displaystyle
y_{ij}^{} \frac{\,v_s^{}v_h^{}M_3^{}\,}{\sqrt{2\,}M_\Phi^2\,},
\ea
\label{eq:Mnu}
\eeq
where the right-handed neutrinos $\nu_{Rj}^{}$
are defined by the following rotation,
\beq
\ba{rl}
{\nu_R}_j^{} &=\,
\dis\fr{1}{\sqrt{2\,}\,} \!\left( R_{1j}^{}\!+ R_{2j}^{} \right),
\\[2mm]
\nu_{sj}^{} &=\,
\dis\fr{1}{\sqrt{2\,}\,} \!\left( R_{1j}^{}\!- R_{2j}^{} \right) ,
\ea
\eeq
which holds for each given flavor index $j$\,.\,
The orthogonal state
$\nu_{sj}^{}$ is $\,\ZZ$ odd. It has no left-handed partner and will remain massless.
Eq.\eqref{eq:Mnu} realizes the Dirac seesaw and
can generate naturally small neutrino masses.
For instance, setting $\,y_{ij}^{}=O(1)$,\,
$\,M_3^{}\!=O(M_\Phi^{})\!=\!O(10^9)$GeV,
and $v_s^{}\!=O(\MeV)$, we obtain $\,m_{\nu ij}^{}=O(0.1)$eV,
which agrees with the current neutrino oscillation data\,\cite{PDG}.

\vspace*{1mm}

Since the SM Higgs boson mass $M_{h}^{}$ is much larger than
the masses of the light scalar $S$ and gauge boson $X^\mu$ as well as the Dirac neutrinos
$(\nu_L^{},\,\nu_R^{})$, it is more convenient to integrate out the SM Higgs doublet $H$
in the low energy effective theory of $S$, $X^\mu$ and $\nu_{L,R}^{}$\,.
The neutrino effective interactions then take the following form,
\beqa
{\cal L} =  -{y'_{ij}}
\bar{\nu}_{Li}^{} \nu_{Rj}^{}S
+\frac{\,g_x^{}}{2}\bar{\nu}_{sj}^{}\gamma^\mu\nu_{Rj}^{}X_\mu^{}
+ \text{h.c.}\,,
\eeqa
where the effective Yukawa coupling
\beqa
\label{eq:y'}
y'_{ij}\,=\,
\frac{\,\sqrt{2\,}m_{\nu ij}^{}\,}{\,v_s^{}\,},
\eeqa
and $\,g_x^{}\,$ is the gauge coupling of $\,U(1)_X^{}$.
Setting the gauge coupling $\,g_x^{}\!=O(0.1)\,$ and the scalar VEV
$v_s^{}\!=\!O(\MeV)$,
we find that the dark photon $X^\mu$ acquires a small mass
via spontaneous symmetry breaking,
$\,m_X^{}\!=g_x^{}v_s^{}\!=O(\keV)\,$.

\vspace*{1mm}

We will demonstrate that the $U(1)_X^{}$ gauge coupling can generate the desired
neutrino self-interaction with scale
$\la S \ra\! = O(\MeV)$\, to resolve the Hubble tension.
As to be shown in the next section, the left-handed
neutrino $\nu_L^{}$ will be converted to $\nu_R^{}$ and $\nu_s^{}$
after the BBN, and the dark photon $X^\mu$ can mediate effective scattering
among $\nu_{L,R}^{}$ and $\nu_s^{}$ before recombination.
The resolution of Hubble tension then requires
$\,v_s^{}\!= O(\MeV)$\, and we choose
$M^{}_\Phi\!= O(M_3^{})=O(10^9\GeV)$
to generate realistic Dirac neutrino masses $\,m_\nu^{}\!=O(0.1\eV)$\,.
With these inputs, the effective Yukawa coupling \eqref{eq:y'} has the size
$\,y'\!=\!\sqrt{2\,}m_\nu^{}/v_s^{}\!=O(10^{-7})$.

\vspace*{1mm}

In summary, our low energy effective theory contains the SM particle content plus
additional new particles, including three light Dirac neutrinos with their
right-handed component $\nu_{Rj}^{}$,
the three right-handed massless fermions $\nu_{sj}^{}$,\,
a massive dark photon $X^\mu$ which mediates the neutrino self-interaction,
and a scalar Higgs boson of $\,\sigma^{}\,$
from the real component of the scalar singlet
$\,S=\fr{1}{\sqrt{2}\,}(\sigma+\text{i}\,\omega)$\,.\,
The three light Dirac neutrinos naturally acquire tiny Dirac masses
$\,m_\nu^{}\!=O(0.1\eV)$\, via the Dirac seesaw mechanism.
In the following analysis, we will ignore the detail of the neutrino
flavor mixing for simplicity. We will also set $\,m_X^{}\!=m_\sigma^{}$\,
for our parameter space, which kinematically forbids the decay channel
$\sigma\!\to\! X^\mu X^\mu$.

\vspace*{2mm}
\section{\large Cosmological\,Evolution\,of\,Interacting\,Right-Handed\,Neutrinos}
\label{sec:3}
\label{sec:RHnu_evol}

In this section, we study qualitatively the evolution of neutrino densities
after the decoupling of electroweak interactions. The key point is that
the scattering of $\nu_L^{}$ only produces a trace amount of $\nu_R^{}$ and $\nu_s^{}$
in the very early Universe because of chirality suppression.
As the Universe cools down, this chirality suppression will be highly reduced.
So the scattering of $\nu_R^{}$ and $\nu_s^{}$
with $\nu_L^{}$ becomes efficient and rapidly converts part of $\nu_L^{}$ into
$\nu_R^{}$ or $\nu_s^{}$\,. The neutrino relic before recombination is a mixture of
$\nu_L^{}$, $\nu_R^{}$ and $\nu_s^{}$\,,\,  which can couple tightly with each other
through the dark photon mediator $X^\mu$
and hence delay the neutrino free-streaming time close to matter-radiation equality.
We will discuss the condition for the evolution and various phenomenological
constraints in this section.

\vspace*{1mm}

Since we only consider the epoch with temperature $\,T\!\gg m_\nu^{}$,\,
the neutrinos are highly relativistic. So for the left-handed neutrino
scattering, we can include the neutrino mass effect
up to its first order via mass-insertion on each incoming state $\nu_L^{}$
of the Feynman diagram.  This induces a chirality-flip suppression factor
$\,m_\nu^{}/\!\sqrt{s}$\, in the scattering amplitude. A derivation of this factor
is given in Appendix\,\ref{sec:chiral_flip}.

\vspace*{1mm}

\begin{figure}
	\centering
	\includegraphics[width=0.6\linewidth]{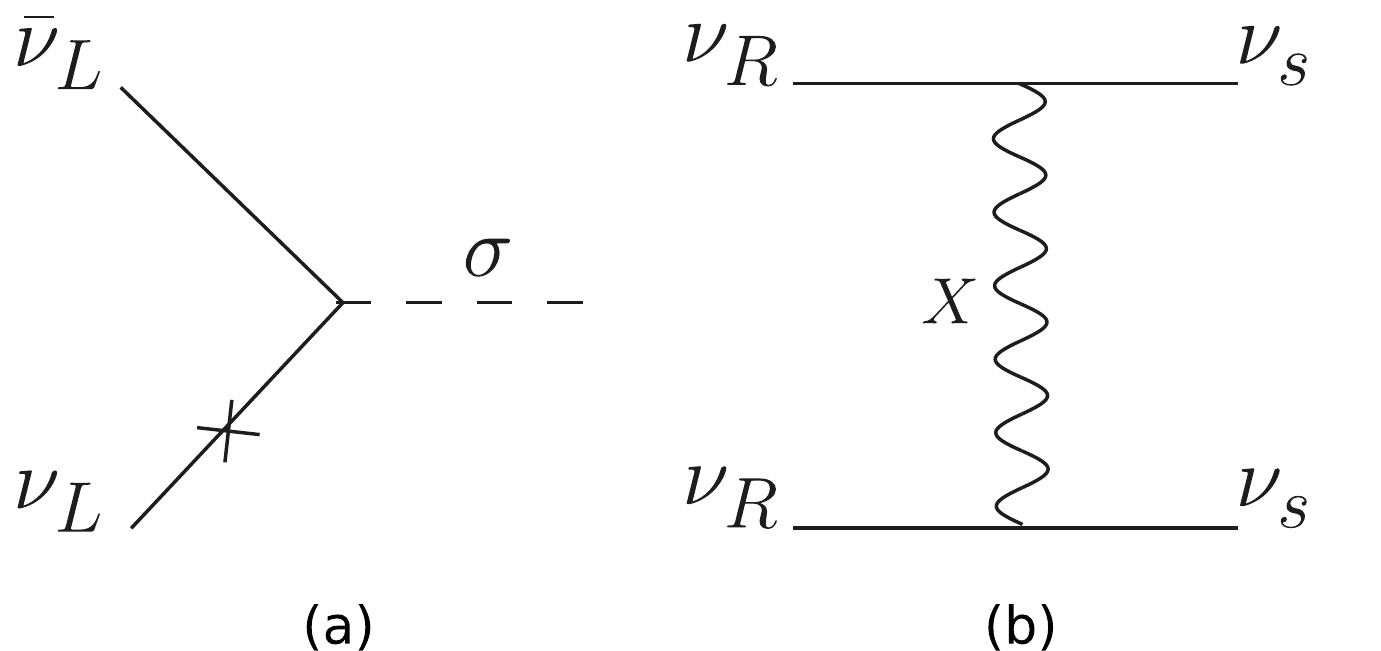}
	\caption{\small
		Panel\,(a):~Inverse decay process $\nu_L^{}\bar{\nu}_L^{}\!\!\to\!\sigma$.
		The produced scalar particle $\sigma$ will decay into
		$\bar{\nu}_L^{}\nu_R^{}$ and $\bar{\nu}_R^{}\nu_L^{}$ subsequently,
		and increase the $\nu_R^{}$ density in the early Universe.
		Panel\,(b):~The conversion process $\nu_R^{}\!\to \nu_s^{}$
		as mediated by the dark photon $X^\mu$.}
	\label{fig:1}
\end{figure}

In the early Universe, only left-handed neutrinos $\nu_L^{}$
are thermalized through electroweak interaction.
After electroweak and $U(1)_X^{}$ symmetry breaking,
$\nu_L^{}$ and $\nu_R^{}$ form massive Dirac particles and oscillate into each other.
The right-handed neutrinos can then be produced out-of-equilibrium via
annihilation process $\,\nu_L^{}\bar\nu_L^{}\!\to\sigma$\,
(incuding a mass insertion of $\,m_\nu^{}\bar\nu_L^{}\nu_R^{}\!+\text{h.c.}$)
as shown in Fig.\,\ref{fig:1}a, with the subsequent $\sigma$ decays
$\,\sigma\!\!\to\!\nu_L^{}\bar{\nu}_R^{},\nu_R^{}\bar{\nu}_L^{}$.%
\footnote{%
	Note that $\nu_R^{}$ can also be produced by $2\!\to\!2$ scattering such as
	$\nu_L^{}\nu_L^{}\!\!\leftrightarrow\!\nu_R^{}\nu_R^{}, \nu_s^{}\nu_s^{}$
	by exchanging a $t$-channel $\sigma$ or $X^\mu$. However, as will be shown below,
	the small amount of $\nu_R^{}$ or $\nu_s^{}$ is only important well after BBN.
	The $2\!\to\!2$ scattering rate at this temperature is much smaller than the
	inverse decay rate by a factor of $\,T^2\!/v_s^2$\,.
	So we will ignore the $2\!\to\!2$ production processes hereafter.}
The thermally averaged cross section of this process can be estimated as
\beqa
\la \sigma v\ra_{\!LL\sigma}^{}
\approx \left( \frac{m_\nu^{}}{m_\sigma^{}} \right)^{\!\!2}
\!\la \sigma v\ra_{\!LR\sigma}^{}\,,
\eeqa
where $\la \sigma v\ra_{LR\sigma}^{}$ is the averaged cross section of
$\,\nu_L^{}\bar{\nu}_R^{}\!\!\to\!\!\sigma$\,
given in Eq.\eqref{eq:ids}.
We see that $\la \sigma v\ra_{LL\sigma}^{}$ is highly suppressed by
$\,m_\nu^4/(m_\sigma^2 v_s^2)\,$,\,
so this annihilation process is extremely slow
and always out of thermal equilibrium in the early Universe.
The produced $\sigma$ bosons then decay predominantly
to $\nu_L^{}\bar{\nu}_R^{}$ and $\nu_R^{}\bar{\nu}_L^{}$,
leading to a net increase of $\nu_R^{}\,(\bar\nu_R^{})$ density.
The small amount of produced $\nu_R^{}$ neutrinos can scatter effectively
among themselves through the $U(1)_X^{}$ gauge interaction.
To see this, we estimate
the density of $\,\nu_R^{}\,$ as
$\,n_{\nu_R^{}}^{} \!\!\!\sim
\,n_{\nu_L^{}}^2 (\frac{m_\nu^{}}{m_\sigma^{}})^{2}
\la \sigma v \ra_{LR\sigma}^{} H^{-1}$,\,
where $n_{\nu_L^{}}^{}$ is the total left-handed neutrino density.%
\footnote{The convention of number density $n_j^{}$ of a particle species
	$\,j$\, in this paper is always defined as the total number density
	including both particles and their antiparticles from all three generations.}
The $\nu_R^{}$ scattering process such as Fig.\,\ref{fig:1}b
will become efficient when
\beqa
H \,\lesssim\,
\tilde\Gamma \approx n_{\nu_R}^{}\!\la \sigma v\ra_{\!RR}^{}
\eeqa
with
$\,\la \sigma v\ra_{\!RR}^{}$\, the characteristic cross section of scattering
between the right-handed particles $\nu_R^{}$ and $\nu_s^{}$ given in Eq.\eqref{eq:sigR}.
This condition is easily satisfied during the interested epoch with $T\lesssim O(\MeV)$.
For the similar reason, the conversion
$\sigma\sigma\!\leftrightarrow\! X^\mu X^\mu$,
$\nu_R^{}\bar{\nu}_s^{}\!\leftrightarrow\! X^\mu$,
and $X^\mu X^\mu\leftrightarrow \nu_R^{}\bar{\nu}_R^{},\,\nu_s^{}\bar{\nu}_s^{}$
are also efficient because they are also induced by the $U(1)_X^{}$ interaction.
Hence, a small amount of tightly coupled fluid $\cal T$ which consists of $\nu_R^{}$,
$\nu_s^{}$, $\sigma$, and $X^\mu$\, is produced from
$\nu_L^{}$ scattering in the early Universe. Since the number of $X^\mu$ particle is evidently violated
by these reactions, the tightly coupled fluid $\cal T$ has a vanishing chemical potential and
a negligible initial temperature.

\begin{figure}
	\centering
	\includegraphics[width=0.6\linewidth]{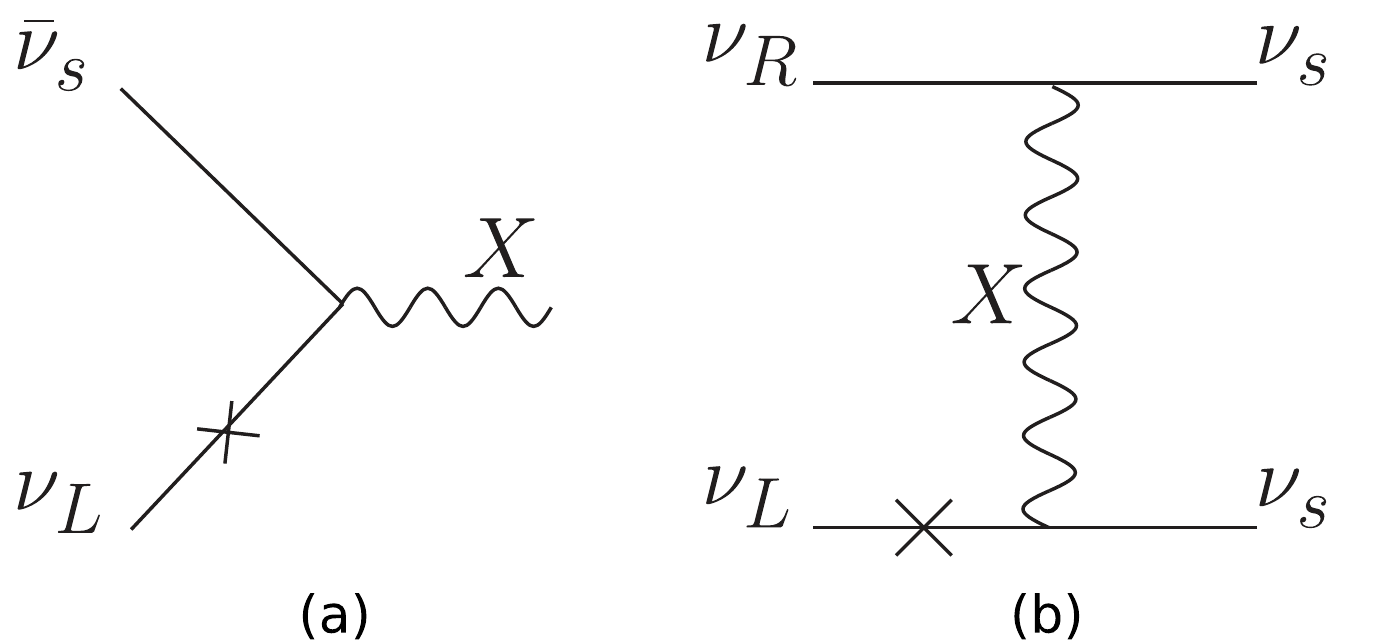}
	\caption{\small
		Conversion processes $\,\nu_L^{}\!\to\!\nu_R^{},\nu_s^{}$\,.
		Panel\,(a):~The inverse decay $\,\nu_L^{}\bar{\nu}_s^{}\!\to\! X^\mu$,\,
		where the final state $X^\mu$ predominantly decays into $\nu_R^{}\bar{\nu}_s^{}$ or
		$\nu_s^{}\bar{\nu}_R^{}$\,.\,
		Panel\,(b):~An example diagram of the conversion process
		$\,\nu_L^{} R\rightarrow\!R R$\,,\,
		where $R$\, denotes $\nu_R^{}$ or $\nu_s^{}$ (or their antiparticles)
		in any family. Our analysis includes all possible channels of
		$X^\mu$ exchanges.}
	\label{fig:Rconv}
	\label{fig:2}
\end{figure}

\vspace*{1mm}

The generated $\,\nu_R^{}\,$ and $\,\nu_s^{}\,$
catalyze the conversion of left-handed neutrinos to
right-handed ones through much faster conversion processes
$\,\nu_L^{}\bar{\nu}_s^{}\!\!\to\!\! X^\mu$\, and
$\,\nu_L^{}R\!\!\to\!\! R R$\,,\, as shown in Fig.\,\ref{fig:2}.
Here $R$ denotes the particle from all three families of
$\nu_R^{}$, $\nu_s^{}$ and their antiparticles.
The cross sections of both conversion processes in
Fig.\,\ref{fig:2} are suppressed by one less factor of $m_\nu^2$
than the annihilation process in Fig.\,\ref{fig:1}(a).
The thermally averaged cross section of
$\,\nu_L^{} \bar{\nu}_s^{}\!\rightarrow\! X^\mu$
and $\,\nu_L^{} R\!\rightarrow\! R R$\,
are given by Eqs.\eqref{eq:idxlr} and \eqref{eq:sigLR},
respectively.
As the Universe cools down,
the conversion rate increases
because the reaction energy is closer to the $X^\mu$ resonance and
the chirality factor $m_\nu^{}/E_\nu^{}$ also becomes larger.
Below a certain temperature $T_c^{}$, the $R+\nu_L^{}$ scattering becomes efficient
and equilibrates their temperature:
\beqa
H \,\lesssim\,\Gamma_\text{conv}^{}\!\equiv\, n_{\nu_L^{}}^{}\!\!
\left( \la \sigma v \ra_{\rm LR} + \la \sigma v \ra_{LRX}^{} \right),
\qquad (\text{for }T\lesssim T_c^{}).
\eeqa
The number of right-handed particles $R$ in a conformal volume
can increase exponentially in a Hubble time
by a factor of $\,\sim\!e^{\Gamma_\text{conv}/H}$\,
via continuously converting $\nu_L^{}$ to $\nu_R^{}$ and $\nu_s^{}$.
This domino effect converts left-handed neutrinos to right-handed neutrinos rapidly,
until $\,n_{\nu_L^{}}^{}\!\!=n_{\nu_R^{}}^{}\!=n_{\nu_s^{}}^{}$,\,
a stationary configuration determined by the principle of detailed balance.
The neutrinos $\nu_L^{}$, $\nu_R^{}$ and $\nu_s^{}$ scatter effectively
with each other, stalking the free-streaming of neutrinos,
which is the key ingredient to shrink down the \,$r_{\ast}^{}$ while
keeping $\,\ell_d^{}$ intact.

\vspace*{1mm}

\begin{figure}[t]
	\centering
	\includegraphics[width=0.75\textwidth]{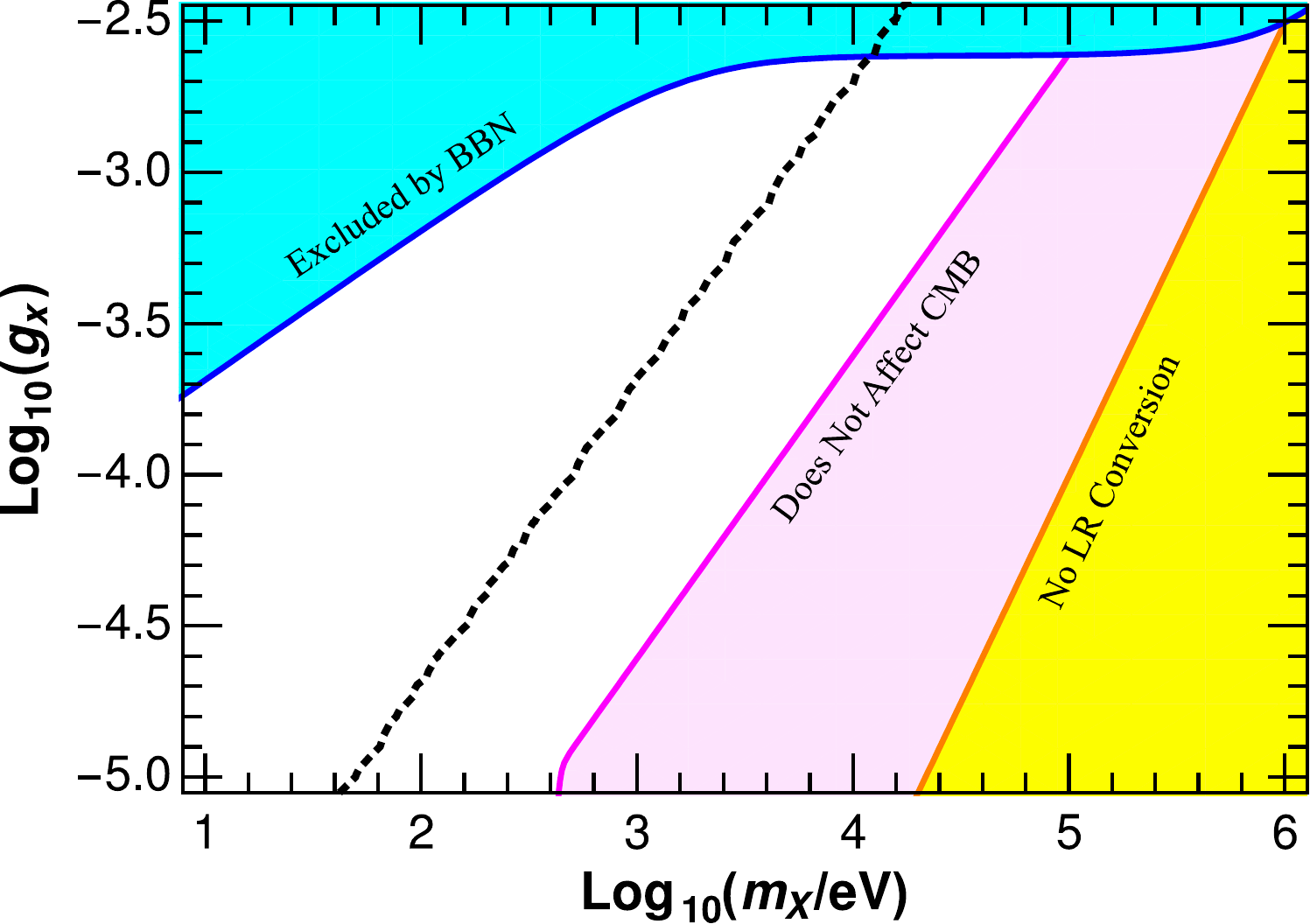}
	\caption{\small
		Constraints on the dark photon gauge coupling $g_x^{}$
		and its mass $m_X^{}$. Here we input a typical neutrino mass
		$m_\nu^{}\!=0.05$\,eV. The blue region overproduces $\nu_R^{}$ and $X^\mu$ before decoupling
		of the neutrino electroweak interaction and is therefore excluded by the BBN.
		In the yellow region, the conversion
		$\,\nu_L^{}\!\!\to\!\nu_R^{}, \nu_s^{}$\,
		is never efficient. In the pink region, neutrinos free-stream too early and
		behave effectively as the SM neutrinos for CMB observation.
		Only the white area is allowed. On the black dotted curve,
		right-handed neutrinos begin free-streaming
		at $\,z\sim 8000$\,,\, which is significantly delayed as compared
		to the standard $\Lambda$CDM model.}
	\label{fig:3}
\end{figure}

There are several conditions that needs to be satisfied by our model.
The rapid conversion process should not happen before decoupling of the
neutrino electroweak interaction at
$\,T\!=\!O(\MeV)$.\, Otherwise, the right-handed neutrinos could be in
equilibrium with the thermal bath and increase the total neutrino density.
This can in turn populate the gauge boson $X^\mu$ of mass
$\,m_X^{}\!=\!O(\keV)$\,
in the early Universe, where $X^\mu$ mediates neutrino self-interaction.
Such an increase of $N_\text{eff}^{}$
was severely constrained by primordial deuterium measurement and tends
to disfavor the self-interacting Dirac neutrinos\,\cite{Blinov:2019gcj}.
But we can avoid this in our model by requiring the total rate
$\Gamma_\text{conv}^{}$ of $\nu_R^{}$
scattering with $\nu_L^{}$ be smaller than the Hubble rate before BBN,
\beqa
\frac{\,\Gamma_\text{conv}^{}\,}{H}\bigg|_\MeV \lesssim 1\,.
\eeqa
This imposes an upper bound on the gauge coupling
$g_x^{}$ for each given dark photon mass $m_X^{}$ as shown for the case of
$m_\nu^{}=0.05\eV$ in the blue shaded region of
Fig.\,\ref{fig:3}. For instance,  Fig.\,\ref{fig:3} gives
$\,g_x^{}\!\lesssim\! 2\!\times\!10^{-4}$\,
for $\,m_{X}^{}\!=\!10\,$eV, and
$\,g_x^{}\!\lesssim 25\!\times\!10^{-4}$\,
for $\,m_{X}^{}\!=\!10^4$\,eV.

\vspace*{1mm}

Since the conversion rate peaks at $\,E\sim m_X^{}$,\,
the following condition should be satisfied as well,
\beqa
\label{eq:Gamma>H(M_X)}
\frac{\,\Gamma_\text{conv}^{}\,}{H}\bigg|_{T=m_X^{}}^{}\gtrsim 1 \,,
\eeqa
so that the rapid conversion process
$\nu_L^{}\rightarrow \nu_{R}^{},\nu_s^{}$
can occur in the early Universe.
This excludes the yellow shaded region in Fig.\,\ref{fig:3}.

\vspace*{1mm}

The resolution of the Hubble tension requires
that the neutrino non-free-streaming alters the damping tail of the CMB power spectrum\,\cite{Kreisch:2019yzn}.
The scattering should be efficient when the relevant Fourier mode corresponding to the damping tail enters the Hubble radius. As a benchmark, the Fourier modes corresponding to multipoles
$\,\ell\!\sim\!2000$\, enter the Hubble radius at
$\,T_{\rm t}^{}\!\sim\! 10\,$eV.
The right-handed neutrino scattering should be efficient
around this epoch and therefore satisfies
\beqa
\frac{\,\Gamma^{}_R\,}{H}\bigg|_{T_{\rm t}}>1\,,
\eeqa
where
$\,\Gamma_R^{}\! = n_{R}^{} (\la\sigma v \ra_{RRX}^{} \!+\! \la\sigma v \ra_{RR}^{})$\,
is the total rate of the $\nu_R^{}\!+\nu_s^{}$ scattering.
By assuming that the left-right handed neutrino conversion
already finished before this epoch,
the detailed balance of the conversion processes such as
$\nu_L^{}\nu_R^{}\LRto \nu_s^{}\nu_s^{}$ and $\nu_R^{}\nu_R^{}\LRto \nu_s^{}\nu_s^{}$
implies
$\,n_R^{}/2\!= n_{\nu_R}^{}\!\!= n_{\nu_s}^{}\!= n_{\nu_L}^{}$.\,
This excludes the pink region in Fig.\,\ref{fig:3}
for the typical neutrino mass value $m_\nu^{}\!=0.05$\,eV
(based on neutrino oscillation data).
Finally, if the free-streaming of neutrinos start too late,
the neutrino self-scattering would strongly alter
the low-$\ell$ part of the CMB power spectrum and thus
deteriorate the fit to the observation\,\cite{Forastieri:2019cuf}.
We note that the recent studies of self-interacting-neutrino
cosmology have a
delayed onset of neutrino free-streaming at $\,z\!\sim\!8000$\,
when modes of $\,\ell\!\approx\! 400$\, enter the Hubble radius~\cite{Kreisch:2019yzn,Lancaster:2017ksf}.
We consider a similar onset time of free-streaming for the right-handed neutrinos
in the current estimate.
(A precise determination of the onset time of free-streaming needs a systematical
fit of the CMB power spectrum which is beyond the current scope.)
As a guideline, we consider the right-handed neutrinos to begin free-streaming
at $\,z\!\sim\! 8000$\,, and plot this case in Fig.\,\ref{fig:3}
as the black dotted curve.

\vspace*{1mm}

Finally, we comment on the laboratory and astrophysical constraints.
Our model conserves lepton number, hence it is not constrained
by the neutrinoless double-beta decay
measurements\,\cite{Agostini:2015nwa}. 
The major laboratory constraints on our model come from meson decays.
The typical neutrino energy in these processes are
\,$E_\nu^{}\!\!\sim\! O(100\MeV)\!>\!v_s^{}$,\,
so the chirality-flip factor of mass insertion
$\,m_\nu^{}/E_\nu^{}\!\sim\!10^{-9}$\,
is much smaller than the effective neutrino Yukawa coupling $\,y'\,$
to the singlet Higgs boson $\sigma$\,,\, where $\,|y'|\!=\!O(10^{-7})$.\,
This means that the left-handed neutrinos $\nu_L^{}$ from meson decays
could emit dark photon $X^\mu$ only after the mass insertion with the
suppression $\,m_\nu^{}/E_\nu^{}$,\, while $\nu_L^{}$ can emit $\sigma$
boson with the effective Yukawa coupling $y'$.\,
Hence, we expect the effective Yukawa coupling $y'$ to receive nontrivial constraint
from meson decays via emitting $\sigma$ bosons.
The strongest constraint from meson decay on a scalar coupling to
neutrinos\,\cite{Pasquini:2015fjv} arises from measuring the light meson decay spectrum
which was used to search for heavy neutrinos\,\cite{Britton:1992xv,Artamonov:2014urb}.
This sets an upper bound $\,|y'|^2\!<\!3.8\!\times\! 10^{-7}$.\,
The neutrino emission of the Supernova\,1987A
\cite{Hirata:1987hu,Bionta:1987qt}
may also be modified by the emission of $\,\sigma\,$ bosons
from the left-handed neutrinos trapped in the core\footnote{%
	The core-collapse process is not well understood, so the resultant bound
	should be considered as an estimate rather than a strict constraint\,\cite{Bar:2019ifz}.}
through the vertex in Fig.\,\ref{fig:1}a.
From the result of \cite{Blinov:2019gcj}, we derive an upper limit
$\,|{y'}|^2 \lesssim (1-12)\!\times\!10^{-5}/(1\!+\!m_\sigma^{}/\keV)$\,.\,
Both constraints are well satisfied since we have smaller Yukawa coupling
$\,|y'|=O(10^{-7})$\, in the current model.

\vspace*{1mm}

In passing, we note that various non-zero $\,\Delta N_\text{eff}^{}$\,
near the epoch of recombination may also help
to reduce the Hubble tension to different levels.
The right-handed neutrinos in our model
are converted from left-handed neutrinos after their decoupling from
the hot plasma of other SM particles and thus do not introduce
$\Delta N_\text{eff}^{}$ before the BBN.
The BBN constraint
$\,\Delta N_\text{eff}^{}\lesssim 0.5$ \cite{Berlin:2019pbq,Blinov:2019gcj}
comes from the model-dependent baryon-to-photon ratio
along with the measured primordial abundance of
$\,Y_{\rm p}^{}$ \cite{Aver:2015iza} and [D/H] \cite{Cooke:2017cwo}.
To realize the model considered in Ref.\,\cite{Kreisch:2019yzn}
with $\,\Delta N_\text{eff}^{}\approx 1$\,,
the additional $\Delta N_\text{eff}^{}$ needs to be generated
in the epoch between BBN and recombination.
Indeed, the decay of the massive particles $\sigma$ and $X^\mu_{}$ as they
decouple from the $\cal T$ fluid at $T\lesssim m_\sigma^{},m_X^{}$ heats up the
neutrinos\cite{Chacko:2003dt,Berlin:2017ftj,Berlin:2018ztp}. The increase in the neutrino temperature
can be estimated by conservation of energy density at the left-right neutrino conversion
and the conservation of entropy density at $X^\mu$, $\sigma$ decoupling:
\be
T_\nu^{}\approx
\left(\!\frac{4}{11}\!\right)^{\!\!\frac{1}{3}}\!
\left(\!\frac{21}{\,79\,}\!\right)^{\!\!\frac{1}{4}}\!
\left(\!\frac{79}{\,63\,}\!\right)^{\!\!\frac{1}{3}}
T_\gamma^{}\,.
\label{eq:tnu_fin}
\ee
This corresponds to $\Delta N_\text{eff}\approx 0.23$.
(In the SM, neutrinos do not decouple instantaneously and
are slightly heated by $e^-e^+$ annihilations, resulting in
$\,\Delta N_\text{eff}^{}\!\approx\! 0.046$ \cite{Mangano:2001iu,Mangano:2005cc}.
We ignore this minor contribution in the current study.)
The remaining $\,\Delta N_\text{eff}^{}$\, required for solving the Hubble tension
can be achieved by entropy injection from the dark sector\,\cite{Alcaniz:2019kah}.
Alternatively, one may assume a smaller $\Delta N_\text{eff}^{}$
that is consistent with the BBN constraint at the expense of less reduction
of the Hubble tension as in the case of \cite{Ghosh:2019tab}.
Since there are many possibilities of choosing $\Delta N_\text{eff}^{}$
which are highly dependent on the dark sector models,
we will focus on the realization of the neutrino self-interaction scenario
as an attractive major resolution in this study.

\vspace*{1mm}

In summary, we have demonstrated in this section that with a suitable choice of
parameter space shown in Fig.\,\ref{fig:3}, the left-handed neutrinos
convert to the right-handed neutrinos only after the BBN.
The final neutrino relic is a mixture of $\nu_L^{}$, $\nu_R^{}$ and $\nu_s^{}$
which scatter with each other before recombination.
The evolution of the neutrino density is consistent with the BBN
and our model is safe under the laboratory and supernovae constraints
on hidden neutrino interactions.

\vspace*{2mm}
\section{\large Evolution of Neutrino Densities by Numerical Analysis}
\label{sec:evolution}
\label{sec:4}

In this section, we will demonstrate the neutrino density evolution
as discussed qualitatively in Section\,\ref{sec:RHnu_evol}.  For this,
we solve the evolution of Boltzmann equation for energy densities numerically for the
$(\nu_L^{},\, \nu_R^{},\, \nu_s^{},\, X^\mu,\, \sigma)$ system
with a given set of parameters.
The numerical result is consistent with
the physical picture given in Section\,\ref{sec:RHnu_evol}.

\vspace*{1mm}

In the parameter space of interest, the decay rate of $X^\mu$
and the scattering rates among $\sigma$,
$\nu_s^{}$, $\nu_R^{}$ and $X^\mu$ are dominated by the $U(1)_X^{}$
gauge interaction and much larger than the Hubble rate
as we have shown in Section\,\ref{sec:RHnu_evol}.
Since the $X^\mu$ number changing reactions is in equilibrium,
we can treat $\nu_s^{}$, $\nu_R^{}$, $\sigma$
and $X^\mu$ as a single tightly coupled fluid $\,\cal T\,$ with
temperature $T_{\cal T}$ and zero chemical potential. The energy density
of the fluid is,
\beqa
\rho_{\cal T}^\text{eq}(T_{\cal T})\,
\equiv\, \rho_{\nu_R^{}}^\text{eq}(T_{\cal T})
+\rho_{\nu_s^{}}^\text{eq}(T_{\cal T})
+\rho_\sigma^\text{eq}(T_{\cal T})
+\rho_X^\text{eq}(T_{\cal T})\,,
\label{eq:nt}
\eeqa
where $\rho^\text{eq}_i(T_{\cal T})$ is the equilibrium density of the particle species $i$
with temperature $T_{\cal T}$ and zero chemical potential,
\beqa
\rho^{\eq}_i(T) \,=\, \frac{g_i^{}}{\,(2\pi)^3\,}\!\int\!\! \text{d}^3 p\,
\frac{E}{\,\exp(E/T)\pm 1\,}\,,
\label{eq:rhoeq}
\eeqa
and $\,g_i^{}\,$ being the degrees of freedom.
$\,\rho_{\nu_R^{}}^\text{eq}$ and $\,\rho_{\nu_s^{}}^\text{eq}$ denote the total
energy densities including both particles and anti-particles summed over the three families.
The Boltzmann equations that govern the evolution of the left-handed neutrino temperature
$T_{\nu_L^{}}^{}$ and the $\cal T$ fluid temperature $T_{\cal T}^{}$ are,
\beqa
\frac{{\rm d} \rho^\text{eq}_{\cal T}(T_{\cal T})}{ {\rm d} t }
+3H\bigg( \rho^\text{eq}_{\cal T}(T_{\cal T}) + P^\text{eq}_{\cal T}(T_{\cal T}) \bigg)
&=& I_E^{\cal T}\,, \\
\frac{{\rm d} \rho^\text{eq}_{\nu_L^{}}(T_{\nu_L^{}}\!)}{ {\rm d} t }
+4H \rho^\text{eq}_{\nu_L^{}}(T_{\nu_L^{}}\!)
&=& I_E^{\nu_L^{}}\,.
\eeqa
$P^\text{eq}_{\cal T}(T_{\cal T})\equiv \sum_i P^\text{eq}_i(T_{\cal T})$ is the total pressure
of each component of the $\cal T$ fluid.
$I_E^{\cal T}$ and $I_E^{\nu_L^{}}$ are the rates of the energy density transferred into the
$\cal T$ fluid and $\nu_L$ respectively.
Their expressions are presented in Eq.\eqref{eq:boltz}.
The fast reactions are contained within the $\cal T$ fluid
and do not appear in $I_E^{\cal T}$ and $I_E^{\nu_L^{}}$.
The processes that evolve $\rho_{\cal T}^\text{eq}(T_{\cal T}^{})$ or
$\rho_{\nu_L^{}}^\text{eq}(T_{\nu_L^{}}^{}\!)$ are those
with at least a mass-insertion suppression
in an external leg or with a Yukawa vertex of coupling $\,y'$
such as those in Fig.\,\ref{fig:1}(a) and Fig.\,\ref{fig:2}.

\vspace*{1mm}

\begin{figure}[t]
	\centering
	\includegraphics[width=0.99\textwidth]{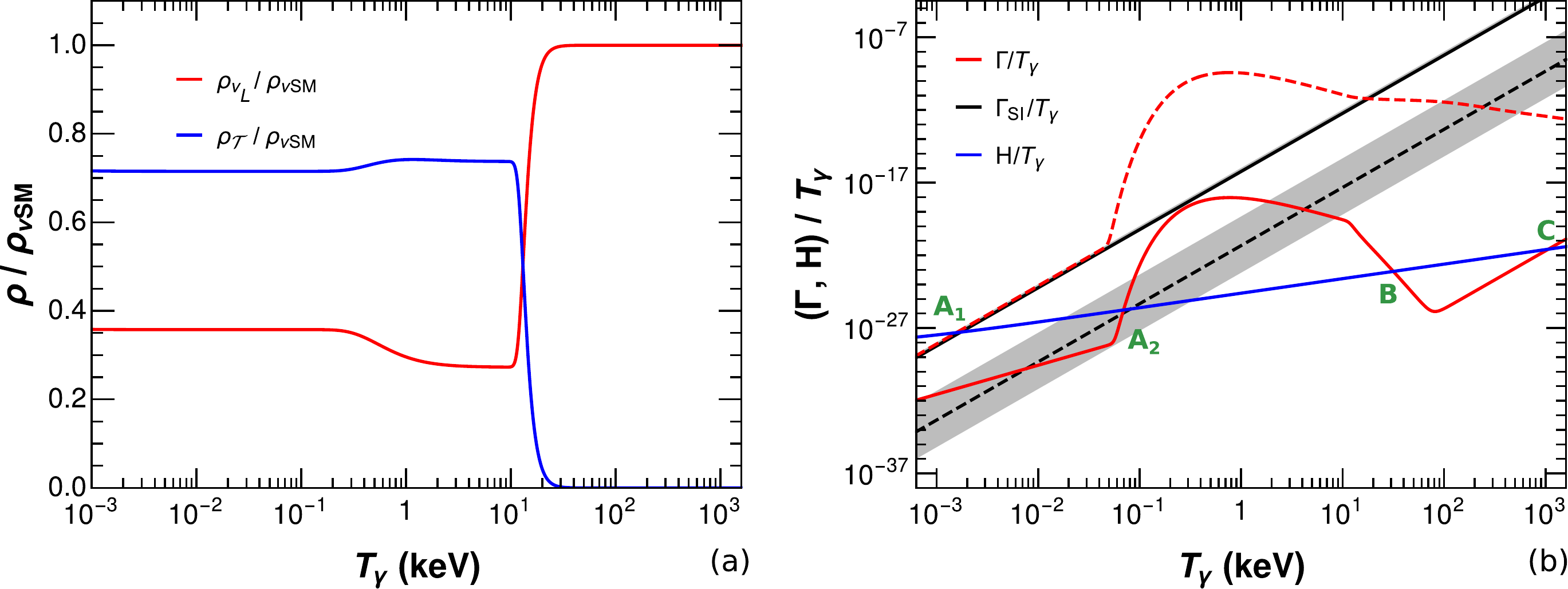}
	\caption{\small
		Panel\,(a):~Evolutions of the scaled energy densities of left-handed
		neutrinos $\rho_{\nu_L^{}}^{}/{\rho_{\nu\text{SM}}^{}}$ (red curve)
		and the tightly-coupled fluid
		$\rho_{\cal T^{}}^{}/{\rho_{\nu\text{SM}}^{}}$ (blue curve) are shown as functions of
		the photon temperature $\,T_\gamma^{}$ (keV), for the interacting Dirac neutrino model
		with $\,m_\nu^{}\!=0.05\,$eV, $v_s^{}\!=6\,\MeV$,\,and $\,m_X^{}\!=1\,\keV$.
		$\rho_{\nu\text{SM}}$ is the Standard Model neutrino energy density
		at given $T_\gamma^{}$.\,
		Panel\,(b):~Scaled reaction rate $\Gamma/T_\nu^{}$
		and Hubble rate $H/T_\gamma^{}$ are plotted
		as functions of $\,T_\gamma^{}$ for our interacting Dirac neutrino model.
		The red solid (dashed) curve shows the interaction rate of
		left-handed (right-handed) Dirac neutrinos with
		$\,m_\nu^{}\!\!=\!0.05$\,eV,\, $v_s^{}\!\!=\!6\,\MeV$, and
		$\,m_X^{}\!\!=\!\!1\,\keV$.\,
		The Hubble rate is shown in the blue curve.
		The black solid (dashed) curve shows the reaction rate of the
		strong (moderate) neutrino self-interaction with
		$\,\text{log}_{10}^{}(G_\text{eff}^{} \MeV^2 )\!=\! -1.35^{+0.12}_{-0.066}\,
		(-3.90^{+1.0}_{-0.93})$ \cite{Kreisch:2019yzn},
		where the shaded gray region presents the 68\% confidence limit in each case
		and the shaded region around the black solid curve is too narrow to be visible.
		Note that the vertical axis of panel\,(a) is plotted in linear scale,
		while it is in log-scale for panel\,(b).
	}
\label{fig:4}
\vspace*{2mm}
\end{figure}

We solve the Boltzmann equations of the
$\nu_L^{}\!\!-\!{\cal T}$ system with the choice of parameters
$\,m_\nu^{}\!=0.05\,\eV$, $v_s^{}=6\,\MeV$,\,
and $\,m_X^{}=m_\sigma^{}=1\,\keV$.\,
Fig.\,\ref{fig:4} presents our results.
In Fig.\,\ref{fig:4}a,
we plot the evolution of $\rho_{\nu_L^{}}^{}/{\rho_{\nu\text{SM}}^{}}$
and $\rho_{\cal T^{}}^{}/{\rho_{\nu\text{SM}}^{}}$\,
as a function of the photon temperature $T_\gamma^{}$\,.
$\rho_{\nu\text{SM}}$ is the Standard Model neutrino energy density.
The generation of the $\cal T$ fluid becomes rather rapid at
the temperature $\,T_\gamma^{}\approx 20\,\keV$, which is much later than the decoupling
of neutrinos from other SM particles.
This is the key feature of the cosmological evolution of our model
as discussed in Section\,\ref{sec:RHnu_evol}.
When the temperature drops below $\sim\!\!100\,\eV$ in Fig.\,\ref{fig:4},
$X^\mu$ and $\sigma$ no longer remain
in the $\cal T$ fluid because of their large masses.
The energy densities of
$\nu_L^{}$, $\nu_s^{}$ and $\nu_R^{}$
are then equal to each other, with
$\,\rho_{\nu_L^{}}^{}\!\!=\rho_{\nu_s^{}}^{}\!\!=\rho_{\nu_R^{}}^{}\!\!
\approx 3.22\times\rho_{\nu\text{SM}}^{}/3$\,. This is close to
the estimation of $\Delta N_\text{eff}\approx 0.23$ corresponds to
Eq.\eqref{eq:tnu_fin}.

\vspace*{1mm}

In Fig.\,\ref{fig:4}(b),
we plot the scaled reaction rate of neutrinos ($\Gamma/T_\gamma^{}$)
according to the evolution of densities in Fig.\,\ref{fig:4}(a).
Here, the interaction rate of left-handed (right-handed) neutrinos
is presented by the red solid (dashed) curve.
The interaction rate of $\nu_s^{}$ also follows the same
red dashed curve as the right-handed neutrino $\nu_R^{}$\,.\,
The Hubble rate is depicted by the blue curve.
The intersections of the scattering rates and the Hubble rate
are distinctive epochs of the cosmological evolution,
and we mark their locations by the bold letters
$\textbf{A}_{1}^{}$, $\textbf{A}_{2}^{}$, $\textbf{B}$ and $\textbf{C}$,
respectively.
The scattering rate of the left-handed neutrino is initially
dominated by the electroweak interaction at the very right side
of the red solid curve. As the Universe cools down, the weak interaction becomes inefficient
at the epoch {\bf C} (the intersection of the red solid curve with the blue curve)
and the neutrinos decouple from the hot plasma.
In contrast, the reaction rate of the small amount of right-handed neutrinos for
$\,T\!\gtrsim 20$\,keV\, is dominated by its scattering within the
$\cal T$ fluid such as the process shown in Fig.\,\ref{fig:1}(b).
As the energy densities $\,n_{\nu_R^{}}^{}$\! and $\,n_{\nu_s^{}}^{}$ increase,
the reaction rate of $\nu_L^{}$ becomes dominated
by its scattering with $\nu_R^{}$ and $\nu_s^{}$,\,
and the conversion becomes rapid at the epoch {\bf B}.
This corresponds to the sharp increase of
$\,\rho_{\cal T}^{}$\, around $\,T\!\!=\!20$\,keV
in Fig.\,\ref{fig:4}(a).
Note that the panels~(a) and (b) in Fig.\,\ref{fig:4} are plotted
in linear and log scale, respectively.
Eventually, the neutrino gas becomes a mixture of
$\nu_L^{}$, $\nu_s^{}$, and $\nu_R^{}$\,.
The reaction rate of $\,\nu_R^{}$\, and $\,\nu_s^{}$\, gets dominated by the
scattering between themselves such as the process
in Fig.\,\ref{fig:1}(b).
Because of the mass insertion,
$\nu_L^{}$ scatters less frequently than other components
as is evident from the difference between the red dashed curve
and the red solid curve. The left-handed neutrinos start to free-stream when
$\,T\!\approx\! 70$\,eV\, at the epoch $\textbf{A}_2^{}$,
much earlier than the right-handed neutrinos which start to free-stream
when $\,T\!\approx\! 1$\,eV at the epoch $\textbf{A}_1^{}$.

\vspace*{1mm}

In comparison with the standard model (SM) neutrinos
which start to free-stream at the epoch {\bf C},
the delayed onset of neutrino free-streaming in our model leads to
phase shifts and amplification of acoustic peaks in the CMB power spectrum;
these effects can be compensated by shifts of other cosmological parameters
that implies larger Hubble constant and $N_\text{eff}$.
The previous study of the CMB power spectrum
suggests\,\cite{Kreisch:2019yzn}
that a larger value of Hubble constant up to
$\,H_0^{}=72.3\pm 1.4\,{\km\,\SEC^{-1} \Mpc^{-1}}$
can be accommodated by the CMB measurements with
$\,\Delta N_\text{eff}^{}\!\approx\! 1$ as long as the active neutrinos scatter with
themselves through an effective interaction
$\,G_\text{eff}^{}\,\bar{\nu}{\nu}\bar{\nu}{\nu}\,
$.\,
This is a fairly  model-independent approach since it does not depend on
details of how this effective interaction arises.
For this, Ref.\,\cite{Kreisch:2019yzn} considered
a ``strongly interacting'' scenario with
$\,\text{log}_{10}^{}(G_\text{eff}^{} \,\MeV^2)\!=\!-1.35^{+0.12}_{-0.066}$\,
and a ``moderately interacting'' scenario with
$\,\text{log}_{10}^{}(G_\text{eff}^{} \,\MeV^2)\!=\!-3.90^{+1.0}_{-0.93}$\,,\,
which accommodate the Hubble constant of values
$\,H_0^{}=72.3\pm 1.4\,{\km\,\SEC^{-1} \Mpc^{-1}}$
and $\,H_0^{}\!=\!71.2\pm 1.3\,{\km\,\SEC^{-1} \Mpc^{-1}}$,\, respectively.
To make use of the fits of \cite{Kreisch:2019yzn},
in Fig.\,\ref{fig:4}(b) we plot as a reference
the reaction rate of the central value of the
strong (moderate) neutrino self-interaction by the black solid (dashed) curve
and the $68\%$ confidence region by the gray shaded region. The gray region
around the black solid curve is too narrow and nearly invisible.
A direct numerical comparison of the reaction rate is given at the end of
Appendix\,\ref{sec:boltz_evolv}.

\vspace*{1mm}

Despite some difference in the detailed form of the neutrino self-interactions,
Fig.\,\ref{fig:4}(b) shows that
the right-handed (left-handed) neutrinos
in our model start to free-stream at roughly the same epoch
$\textbf{A}_1^{}$ ($\textbf{A}_2^{}$) as
the reference scenario of the strongly (moderately) self-interacting neutrinos.
The reaction rate of right-handed neutrinos traces closely that of the
strongly self-interacting neutrinos after $\textbf{A}_1^{}$
so their impacts on the CMB power spectrum are mainly the same.
On the other hand, the reaction rate of the left-handed neutrinos in our model has a different shape
from that of the moderately self-interacting neutrinos
after $\textbf{A}_2^{}$ and implies different effects on
the high-$\ell$ tail of the CMB power spectrum.
We note that the red solid curve lies entirely within the
$68\%$ confidence region of the black dashed curve so
it can be viewed as an interpolation of moderately self-interacting scenario of
various interaction strength within the 68\% confidence level. A more careful study
of the CMB power spectrum is desirable to pin down the exact impact from the
temperature dependence the left-handed neutrino opacity.
Another interesting aspect of the self-interacting neutrino cosmology is that it
allows a larger $\sum\! m_\nu^{}$ than the $\Lambda$CDM model.
Increasing the neutrino masses raises the reaction rate for the left-handed neutrinos
through the mass-insertion factor,
while keeping the rates of the right-handed neutrinos unchanged.
This can potentially alleviate the
Hubble tension through a stronger impact on the high-$\ell$ tail.

\vspace*{1mm}

With a composition of $2/3$ strongly interacting right-handed neutrinos and
$1/3$ left-handed moderately interacting neutrinos, the cosmic neutrino relic
in our model can be regarded as
an interpolation of the two scenarios of strongly and moderately interacting neutrinos
in Ref.\,\cite{Kreisch:2019yzn}. Assuming entropy injection of
$\,\Delta N_\text{eff}^{}\!\simeq\! 1$\, from the dark sector after the BBN
as we discussed in Sections\,\ref{sec:RHnu_evol},
the inferred Hubble constant from CMB observation of our model should lie between
those of the strongly and moderately interacting regimes.
Therefore, the late onset of neutrino free-streaming in our model
is expected to be consistent with a larger Hubble constant in the range of
$\,H_0^{}\!\simeq\! (70 \!-\! 72)\,{\km\,\SEC^{-1}\Mpc^{-1}}$
without deteriorating the fit to CMB observations.
In this way, our model can mainly remove
the tension with the local measurements of Hubble constant
$\,H_0^{}=74.0\pm 1.4\,{\km\,\SEC^{-1} \Mpc^{-1}}$ \cite{Riess:2019cxk}.

\vspace*{1mm}

A direct numerical comparison of the reaction rates with
Ref.\,\cite{Kreisch:2019yzn}
is given at the end of our Appendix\,\ref{sec:boltz_evolv}. We find that
with the choice of parameters
$m_\nu^{}\!=0.05\,\eV$ and $v_s^{}\!=6\,\MeV$,\,
the right-handed and left-handed neutrinos interact with the neutrino gas,
which have effective 4-neutrino couplings
$\,G_\text{R}^{}\!=\! 10^{-1.28}\MeV^{-2}$\, and
$\,G_\text{L}^{}\!=\, 10^{-4.01}\MeV^{-2}$,\, respectively,
at $T\!\approx\! 10\,\eV$.
Hence, the right-handed neutrinos $\nu_R^{}$ and $\nu_s^{}$
(which make up $2/3$ of the neutrino gas in our model) behave like
the strongly self-interacting neutrinos of Ref.\,\cite{Kreisch:2019yzn} with
$\,\text{log}_{10}^{}(G_\text{eff}^{} \,\MeV^2)\!=\!-1.35^{+0.12}_{-0.066}$\,,
while the left-handed neutrinos $\nu_L^{}$ (which make up the remaining $1/3$\,
of the neutrino gas in our model) behave like the
moderately self-interacting neutrinos with
$\,\text{log}_{10}^{}(G_\text{eff}^{} \,\MeV^2)\!=\!-3.90^{+1.0}_{-0.93}$
\cite{Kreisch:2019yzn}.
This is in accordance with our discussion of Fig.\,\ref{fig:4}(b).

\vspace*{1mm}

The above numerical analysis demonstrates that the evolutions
of both the number density and reaction rate are consistent with the
physical picture of Section\,\ref{sec:RHnu_evol}.
Hence, we find that our current Dirac seesaw model provides a viable resolution
to the Hubble tension problem, with wide parameter space shown
in Fig.\,\ref{fig:3}.

\vspace*{2mm}
\section{\large Conclusions}
\label{sec:conclusion}
\label{sec:5}

The discrepancy of the Hubble constant measurements concerns the
cosmological observations inferred from the early and late Universe,
and is fairly robust, ranging from $4\sigma$ to $6\sigma$
deviations\,\cite{Verde:2019ivm}.
If this tension persists, it will point to new physics in
the dynamics of the cosmological expansion,
beyond the standard $\Lambda$CDM cosmology.
Such new physics resolution could arise from the exciting interface
of particle physics and cosmology.
In this work, we proposed a new realization of the self-interacting neutrinos
via Dirac seesaw to achieve the mechanism of shrinking down the physical size of the sound horizon at the last scattering surface,
while keeping the projected Silk damping scale intact.

\vspace*{1mm}

In Section\,\ref{sec:Int_Dirac_nu}, we presented a new Dirac seesaw model
with an anomaly-free dark $U(1)_X^{}$ gauge group,
in which 
the light-dark-photon serves as the mediator and
couples only to the right-handed components of Dirac neutrinos.
It naturally generates small masses for Dirac neutrinos and induces
effective self-interaction for the right-handed neutrinos.
We did not assume the flavor structure for neutrino self-interactions,
unlike the models in the previous literature\,\cite{Blinov:2019gcj}.
Our model can evade both the cosmological and laboratory constraints
because the coupling between the left-handed neutrinos and the dark photon mediator is extremely weak in high energy processes due to the chirality-flip
suppression factor $\,m_\nu^{}/E_\nu^{}$ (Fig.\,\ref{fig:2}).

\vspace*{1mm}

In Sections\,\ref{sec:RHnu_evol} and \ref{sec:evolution}, we studied the cosmological evolution of the
left/right-handed neutrinos, which has nontrivial behaviour because of
the gauge interactions mediated by the dark photon.
We first presented the estimates in Section\,\ref{sec:RHnu_evol},
and the constraints on the dark photon parameter space in Fig.\,\ref{fig:3}.
We then performed numerical analysis
of evolving Boltzmann equations of neutrino densities in Section\,\ref{sec:evolution}.
We demonstrated that after the neutrino decoupling and
for a proper choice of the mediator mass and coupling,
part of the left-handed neutrinos converts into right-handed particles
$\nu_R^{}$ and $\nu_s^{}$ in a very short epoch between the BBN and recombination.
The conversion occurs much later than the neutrino decoupling from the hot plasma
of other SM particles,
so it does not generate extra $\Delta N_\text{eff}^{}$
which would violate the BBN bound.
The right-handed particles are more reactive and couple tightly
to the left-handed neutrinos. The resultant non-free-streaming neutrinos
$\nu_L^{}$, $\nu_R^{}$ and $\nu_s^{}$ cause phase shifts and amplification
of acoustic peaks in the CMB power spectrum, which is a key ingredient
of the resolution to the Hubble tension. Our findings are presented in
Fig.\,\ref{fig:4}.
Setting entropy injection of $\,\Delta N_\text{eff}^{}\!\simeq\! 1$\, after the BBN,
we found that the cosmic neutrino relic in our model can be viewed
as a mixture of strongly-interacting and
moderately-interacting neutrinos\,\cite{Kreisch:2019yzn}.
It is consistent with a larger Hubble constant up to
$\,H_0^{}\!\simeq\! (70\! -\! 72)\,{\km\,\SEC^{-1}\Mpc^{-1}}$
without deteriorating the fit to CMB data.
This mechanism reduces the $\Hz$ discrepancy down to $1\sigma$ level
and thus mainly resolves the Hubble tension by our new scenario
of self-interacting neutrinos.

\vspace*{1mm}

Finally, we note that the left-handed neutrinos and right-handed neutrinos scatter
at different rates in the early Universe and it provides an important target
for the analysis of self-interacting neutrino cosmology.
The hidden $U(1)_X^{}$ interaction is currently unconstrained
by laboratory experiments, but on the other hand
it signifies the role of the CMB observation
to probe non-standard neutrino interactions.
The hidden neutrino interaction may leave a trace on the cosmic neutrino background
where the neutrino energy is extremely small and the chirality-flip factor
is no longer a suppression.

\vspace*{6mm}
\noindent
{\large\bf Acknowledgments}\\
We thank Francis-Yan Cyr-Racine for useful discussions
on Ref.\,\cite{Kreisch:2019yzn}.
JZ thanks Shaofeng Ge for an early discussion.
HJH and JZ were supported in part by the NSF of China 
(under grants No.\,11835005 and No.\,11675086),
and also by the CAS Center for Excellence in Particle Physics (CCEPP),
the National Key R\&D Program of China (under grant No.\,2017YFA0402204),
the Key Laboratory for Particle Physics, Astrophysics and Cosmology
(Ministry of Education), and the Office of Science \& Technology, 
Shanghai Municipal Government (under grant No.\,16DZ2260200).
YZM acknowledges the support of NRF-120385, NRF-120378, NRF-109577, 
and NSFC-11828301.

\appendix

%
%
\newpage

\section{\large Neutrino Chirality Flip via Mass-Insertion}
\label{A:A}
\label{sec:chiral_flip}

In this appendix, we re-derive the chirality-flip factor via mass-insertion
for clarity and completeness.
For each insertion of neutrino mass, an external line of the left-handed neutrino
in the Feynman diagram is modified as
\beqa
\chi_-^{}(p)\to \frac{\,m_\nu^{} p_\mu^{}\bar{\sigma}^\mu\,}{p^2}\chi_-^{}(p)~,
\eeqa
where $\chi_-^{}$ is the left-handed 2-component spinor eigenfunction.
For simplicity, we choose the reference frame such that the direction of the
neutrino momentum $\,\vec{p}\,$ is along $+\hat{z}$\,,\,  and
$\chi_-^{}(p)\!=\!\sqrt{p^0\!+p^3}\bp 0\\1 \ep$.\,
In the limit of $\,p^0\!\!\to\! p^3$\,,\,
the pole factor $(p^0\!-p^3)$ in the denominator will be cancelled
by that in the numerator, and thus we have
\beqa
\frac{~m_\nu^{} p_\mu^{}\bar{\sigma}^\mu\,}{p^2}\!\bp 0\\1 \ep
\rightarrow \frac{m_\nu}{\,2p^0\,}\bp 0\\1 \ep .
\eeqa
This result is also evident from the massive 4-component Dirac spinor:
the right-handed component of a spin-down fermion moving along
$+\hat{z}$ contains a factor ${m_\nu}^{}/(2p^0)$
in the leading order of $m_\nu^{}$.\,
So each mass-insertion in the external line leads to
a factor $\,m_\nu^{}/\!\sqrt{s}\,$ in the amplitude.

\vspace*{2mm}
\section{\large Boltzmann Equations and Cross Sections}
\label{sec:boltz_evolv}.
\label{A:B}
\vspace*{-4mm}

In this appendix, we give the relevant Boltzmann equations
and the thermally averaged cross sections used
to evolve the energy density and neutrino temperatures in Fig.\,\ref{fig:4}.
As we described in Section\,\ref{sec:evolution}, we treat $\sigma$,
$\nu_s^{}$, $\nu_R^{}$ and $X^\mu$ as a single fluid $\cal T$
with temperature $T_{\cal T}$ and zero chemical potential. The evolution of
the neutrino temperatures and densities are then governed by the following equations,
\beqs
\vspace*{-1mm}
\beqa
\frac{{\rm d} \rho^\text{eq}_{\nu_L^{}}(T_{\nu_L^{}}\!)}{ {\rm d} t }
&+& 4H \rho^\text{eq}_{\nu_L^{}}(T_{\nu_L^{}}\!)
\approx
\nonumber \\
&&
- \fr{1}{2}E^L_\sigma n_{\nu_L^{}}^\text{eq} n_R^\text{eq} \la \sigma v \ra_{LR\sigma}^{}
- \fr{1}{2}E^L_X n_{\nu_L^{}}^\text{eq} n_R^\text{eq} \la \sigma v \ra_{LRX}^{}
- E^L_{\nu}n_{\nu_L^{}}^\text{eq} n_R^\text{eq} \la \sigma v \ra_{LR}^{}
\nonumber \\
\label{eq:nulevo}
\hspace*{18mm}
&& 
+ \fr{1}{2} E^{\cal T}_\nu {n_R^{\text{eq}}}^2 \la \sigma v \ra_{LR}^{}
+ \fr{1}{2}E^{\cal T}_X n_X^\text{eq} \Gamma_{XLR}^{}
+ \fr{1}{2}E^{\cal T}_\sigma n_\sigma^\text{eq} \Gamma_\sigma^{} \,,
\\[1.5mm]
\frac{{\rm d} \rho^\text{eq}_{\cal T}(T_{\cal T})}{ {\rm d} t }
&+& 3H\bigg( \rho^\text{eq}_{\cal T}(T_{\cal T}) + P^\text{eq}_{\cal T}(T_{\cal T}) \bigg)
\approx
\nonumber \\
&&
- \fr{1}{2} E^{\cal T}_\nu {n_R^\text{eq}}^2 \la \sigma v \ra_{LR}^{}
- \fr{1}{2} E^{\cal T}_X n_X^\text{eq} \Gamma_{XLR}^{}
- \fr{1}{2} E^{\cal T}_\sigma n_\sigma^\text{eq} \Gamma_\sigma^{}
\nonumber\\
&&  
+ \fr{1}{2} E^L_\sigma \,n_{\nu_L^{}}^\text{eq}\! n_R^\text{eq} \la \sigma v \ra_{LR\sigma}^{}
+ \fr{1}{2} E^L_X n_{\nu_L^{}}^\text{eq}\!n_R^\text{eq} \la \sigma v \ra_{LRX}^{}
+  E^L_\nu n_{\nu_L^{}}^\text{eq}\! n_R^\text{eq} \la \sigma v \ra_{LR}^{}
\nonumber\\
&&
+ E^L_\nu\,{n_{\nu_L^{}}^\text{eq}}^2 \!\left(\! \frac{m_\nu^{}}{\,m_\sigma^{}\,} \!\right)^{\!\!2}\!
\la \sigma v \ra_{LR\sigma}^{} \,,
\eeqa
\label{eq:boltz}
\eeqs
\hspace*{-3mm}
where $\rho^\text{eq}_i(T)$ is the equilibrium density of the fluid $i$
with temperature $T$ and zero chemical potential given by Eq.\eqref{eq:nt}
and \eqref{eq:rhoeq}.
$P^\text{eq}_{\cal T}\equiv \sum_i P^\text{eq}_i$\, is the total pressure
of each component of the $\,\cal T$\, fluid,
\begin{equation}
	P^{\eq}_i(T) \,=\, \frac{g_i^{}}{\,(2\pi)^3\,}\!\int\!\! \text{d}^3 p\,
	\frac{p^2}{3E}\frac{1}{\,\exp(E/T)\pm 1\,}\,,
	\label{eq:rhoeq}
\end{equation}
with $g_i^{}$ the corresponding degrees of freedom.
$n^\text{eq}_i$ js the equilibrium number density of a given type of
particles,
\begin{equation}
	n^\text{eq}_i(T) \,\equiv\, \frac{g_i}{\,(2\pi)^3\,}
	\!\int\!\!\frac{\text{d}^3p }{\,\text{exp}(E/T)\pm 1\,}\,.
\end{equation}
In Eq.\eqref{eq:boltz}, the equilibrium number density for $\nu_L^{}$ is always evaluated
at $T_{\nu_L^{}}^{}$ while those for $\nu_R^{},\,\nu_s^{},\,\sigma$ and $X^\mu$ are always
evaluated at $T_{\cal T}^{}$.
Here we recall that in our notation, each number density $n_i^{}$ contains
both the particles and anti-particles from all families. We approximate the average energy
transferred for each collision as
\be
E^\alpha_i\approx T_\alpha^{}+m_i^{}\,,
\ee
with $i=\sigma,\,X,\,\nu$ and $\alpha=\nu_L,\,{\cal T}$. This is a very crude approximation and
only captures the qualitative behavior at the low and the high temperature limit. Yet, it
suffices to demonstrate that the $\nu_L^{}\rightarrow {\cal T}$ conversion happens sometime
after the BBN. Using a more precise expression leads to an $O(1)$ change in the conversion time.
At the epoch of matter-radiation equality, physical quantities such as the neutrino temperature are mostly
independent of the approximation made on $E^\alpha_i$. We evaluate the thermally averaged
cross sections and decay rates in the Boltzmann equations at the same temperature $T_{\alpha}$
as $E^\alpha_i$ in the same term.
This is a good approximation even for scattering processes between
$\nu_L^{}$ and $\cal T$ that start at very different temperatures. For example, when
$T_{\nu_L^{}}^{} \gg T_{\cal T}^{}$, the center of mass energy of a ${\nu_L^{}}^{} - {\nu_R}^{}$
scattering process would be $O(T_{\nu_L^{}}^{})$. Thus, it is reasonable to evaluate both
$\la \sigma v \ra_{LR}^{}$ and the energy transferred at $T_{\nu_L^{}}^{}$.
Finally, we note that the reduction of $\rho_{\nu_L^{}}$ by the extremely slow
process $\nu_L^{}\bar{\nu}_L^{}\rightarrow\sigma$ is negligible at any time
so we ignore it in Eq.\eqref{eq:nulevo}.

\vspace*{1mm}

For completeness, we also provide the thermally averaged cross sections
used in the Boltzmann equation.%
\footnote{For instance, see Refs.\,\cite{Gondolo:1990dk,Srednicki:1988ce}
	for the method of computing thermal averaged cross sections.}
We compute the thermally averaged decay rates as follows,
\beqs
\begin{align}
	\Gamma_{XLR}^{} &=\,
	\frac{\,3 m_\nu^2 m_X^{}\,}{8v_s^2\pi}\frac{K_1(m_X^{}/T)}{\,K_2(m_X^{}/T)\,}\,,
	\\[1mm]
	\Gamma_\sigma^{} &=\,
	\frac{\,3 m_\nu^2 m_\sigma^{}\,}{8v_s^2\pi}\frac{K_1(m_\sigma^{}/T)}{\,K_2(m_\sigma^{}/T)\,}\,,
	\\[1mm]
	\Gamma_X^{} &=\,
	\frac{m_X^3}{\,4v_s^2\pi\,}\frac{K_1(m_X^{}/T)}{K_2(\,m_X^{}/T)\,}\,.
	\label{eq:decay}
\end{align}
\eeqs
$\Gamma_X^{}$ and $\Gamma_\sigma^{}$ are the thermally averaged decay rate of
$X^\mu$ and $\sigma$, respectively.
$\Gamma_{XLR}^{}$ is the partial decay rate of $X^\mu$ to
$\nu_L^{}\bar{\nu}_s^{},~ \nu_s^{}\bar{\nu}_L^{}$.
Note that for simplicity we have set $m_\sigma^{}=m_X^{}$
and thus the decay mode $\,\sigma\!\to\! X^\mu X^\mu$\, is forbidden.

\vspace*{1mm}

The thermally averaged inverse decay rates are obtained
from the decay rates by the principle of detailed balance,
\beqs
\begin{align}
	\la \sigma v \ra_{RRX}^{} &=\, \frac{\,2n_X^\text{eq}\,}{{n_R^\text{eq}}^2} \Gamma_X^{} \,,
	\label{eq:idx}
	\\
	\la \sigma v \ra_{LRX}^{} &=\,
	\frac{\,n_X^\text{eq}\,}{\,n_R^\text{eq} n_{\nu_L}^\text{eq}\,}
	\Gamma_{XLR}^{} \,,
	\label{eq:idxlr}
	\\
	\la \sigma v \ra_{LR\sigma}^{} &=\,
	\frac{\,n_\sigma^\text{eq}\,}{\,n_R^\text{eq} n_{\nu_L}^\text{eq}\,}
	\Gamma_\sigma^{} \,,
	\label{eq:ids}
\end{align}
\eeqs
where
$\la \sigma v \ra_{\rm RRX}^{}$
, $\la \sigma v \ra_{\rm LRX}^{}$
and $\la \sigma v \ra_{LR\sigma}^{}$ are the thermally averaged cross sections
of $\nu_R^{} R \rightarrow X^\mu$, $\nu_L^{} R \rightarrow X^\mu$
and $\nu_L^{} R \rightarrow \sigma$,\, respectively.
Here $R$ denotes any particle from any family of
$\nu_R^{}$, $\nu_s^{}$ or their antiparticles.
The cross section computed is the average over all possible choices of $R$.\,

\vspace*{1mm}

Finally, the thermally averaged cross sections for the
$2\rightarrow2$ scattering processes are
\beqs
\label{eq:sigRR-LR}
\beqa
\la \sigma v \ra_{RR}^{}
\,&=& \begin{cases}
	\dis\frac{35 T^2}{\,6\pi v_s^4\,}  ,&~ \hspace*{6mm} (T\ll m_X^{}),
	\vspace{0.1cm}
	\\[1mm]
	\dis\frac{m_X^2 }{\,4\pi v_s^4\,}  ,&~ \hspace*{6mm} (T\gg m_X^{}),
\end{cases}
\label{eq:sigR}
%
\\[1mm]
\la \sigma v \ra_{LR}^{}
\,&=& \begin{cases}
	\dis\frac{35 m_\nu^2}{\,144\pi v_s^4\,}  ,&~ (T\ll m_X^{}),
	\vspace{0.1cm}
	\\[1mm]
	\dis\frac{m_\nu^2 m_X^2}{\,32\pi v_s^4 T^2\,}  ,&~ (T\gg m_X^{}),
\end{cases}
\label{eq:sigLR}
\eeqa
\eeqs
where
$\la \sigma v \ra_{LR}^{}$ and $\la \sigma v \ra_{RR}^{}$
are the thermally averaged cross sections of processes
$\,\nu_L^{} R\!\to\! R R$\, and $\,\nu_R^{} R\!\to\! R R$\,, respectively.
We have computed all possible channels of $X^\mu$ exchanges in the
actual analysis.
For Eq.\eqref{eq:sigRR-LR}, we have averaged over all possible choices of $\,R\,$
from any family of $\nu_R^{}$ and $\nu_s^{}$\,, or their antiparticles.
For simplicity, we approximate
the cross sections as piece-wise functions of their limits of
$\,T\!\ll\! m_{X}^{}$\, and $\,T\!\gg\! m_{X}^{}$\,.\,
The scattering near the resonance $\,T\!\sim\!  m_{X}^{}$\,
is mainly captured by the inverse decay cross sections in Eq.\eqref{eq:ids}.
This approximation leads to discontinuities in the slopes of the red curves
in Fig.\,\ref{fig:4}b.

\vspace*{1mm}

For comparison, we also compute the thermally averaged cross section
of the neutrino self-interaction in Ref.\,\cite{Kreisch:2019yzn}.
In this fairly model-independent study,
the active neutrino interaction is parameterized 
by the squared amplitude,\footnote{%
We thank Francis-Yan Cyr-Racine for explaining the convention of
Ref.\,\cite{Kreisch:2019yzn} via email correspondence.}
\beqa
|{\cal M}|^2_{\nu_i} \,\equiv\,
\sum_\text{spin}\sum_{j,k,\ell} 
|{\cal M}|^2_{\nu_i^{}+\nu_j^{}\to\nu_k^{}+\nu_{\ell}^{}} 
\!=
\, 2\,G_\text{eff}^2(s^2\!+t^2\!+u^2) \,.
\label{eq:MKreisch}
\eeqa
This leads to a thermally averaged cross section,
\beqa
\frac{1}{4}\sum_\text{spin}\la \sigma v \ra_\text{SI}^{} = \frac{11}{4\pi}G_\text{eff}^2 \,T^2\,.
\eeqa
Then, we can compare directly the scattering rate $\Gamma_\text{SI}^{}$
of the neutrino self-interaction
in Ref.\,\cite{Kreisch:2019yzn} to the rates of
the right-handed neutrinos $\Gamma_\text{R}^{}$ and the left-handed neutrinos
$\Gamma_\text{L}^{}$ in our model for $\,T\ll m_X^{}$\,,
\begin{subequations}
	\begin{align}
		\frac{\,\Gamma_\text{SI}^{}\,}{\Gamma_R^{}}
		&\,=\, \frac{ \frac{1}{4}\sum\limits_\text{ spin} \la \sigma v \ra_{SI}^{} n_{\nu_{j}^{}}^\prime}
		{~\la\sigma v\ra_{RR}^{} (n_{\nu_R^{}}^{} \!\!+ n_{\nu_s^{}}^{})~}
		\,\equiv\, \frac{\,G_\text{eff}^2\,}{G_\text{R}^2} \,,
		\label{eq:ratioR}
\\[3mm]
		\frac{\,\Gamma_\text{SI}^{}\,}{\Gamma_L^{}}
		&\,=\, \frac{\frac{1}{4}\sum\limits_\text{ spin} \la \sigma v \ra_{SI}^{} n_{\nu_{j}^{}}^\prime}
		{~\la\sigma v \ra_{LR}^{} (n_{\nu_R^{}}^{} \!\!+ n_{\nu_s^{}}^{})~}
		\,\equiv\, \frac{\,G_\text{eff}^2\,}{G_\text{L}^2} \,.
		\label{eq:ratioL}
	\end{align}
\end{subequations}
Here we have defined the effective coupling constants
$\,G_\text{L}^{}$\, and $\,G_\text{R}^{}$\, for direct comparison with
the $\,G_\text{eff}^{}$\, in \eqref{eq:MKreisch}
from Ref.\,\cite{Kreisch:2019yzn},
\begin{subequations}
	\begin{align}
		G_\text{R}^{} & \approx\,3.54\times\!\frac{1}{\,v_s^2\,}\,,
		\\[1mm]
		G_\text{L}^{} & \approx\,0.70\times\!\frac{\,m_\nu^{}\,}{T_{\gamma}}\frac{1}{\,v_s^2\,}\,.
	\end{align}
\end{subequations}
Note that in deriving these relations,
$\,(n_{\nu_R^{}}^{}\!\! + n_{\nu_s^{}}^{})\!\propto 3\times 3\times 2\,T_\nu^3$\, 
in our model,
with $\,T_\nu^{}$\, the neutrino temperature related to the photon temperature by Eq.\eqref{eq:tnu_fin}.
Ref.\,\cite{Kreisch:2019yzn} considered
the scattering between Majorana neutrinos and
$\,n_{\nu_{j}^{}}^\prime\!\!\propto 2\,{{T_\nu}^\prime}^{3}$\, 
is the neutrino number density for each flavor.
$T_\nu^{\,\prime}\!=\!(4/11)^{{1}/{3}}T_\gamma^{}$ as in the standard case.
For $\,m_\nu^{}\!=0.05\,\eV$\, and \,$v_s^{}\!=6\,\MeV$
as we choose in Sec.\ref{sec:evolution}, we obtain
$\,G_\text{R}^{}\!=\! 10^{-1.28}\,\MeV^{-2}$\, and
$\,G_\text{L}^{}\!\!=\! 10^{-4.01}\,\MeV^{-2}$\, at $\,T\!\approx\! 10\,\eV$.
Hence, the right-handed and left-handed neutrinos in our model behave like
the strongly and moderately self-interacting neutrinos of
Ref.\,\cite{Kreisch:2019yzn} with the effective coupling
$\,\text{log}_{10}^{}(G_\text{eff}^{}\,\MeV^2)=-1.35^{+0.12}_{-0.066}$\,
and
$\,\text{log}_{10}^{}(G_\text{eff}^{} \,\MeV^2)=-3.90^{+1.0}_{-0.93}$\,,
respectively.

\end{document}